\documentclass[onecolumn]{aa}

\usepackage[varg]{txfonts}
\include{bookdefs}
\usepackage{graphicx}
\usepackage{xspace}
\usepackage{amsmath}
\usepackage{placeins}
\usepackage{gensymb}
\usepackage{color}
\usepackage{float}

\renewcommand{\vec}[1]{\mbox{\boldmath$#1$}}
\newcommand{\music}{\texttt{MUSIC}\xspace}

\definecolor{orange}{rgb}{.9,.3,0}

\begin{document}
\title{Spherical-shell boundaries for two-dimensional compressible convection in a star}
\author{J. Pratt\inst{\ref{inst1}}  \and I. Baraffe\inst{\ref{inst1},\ref{inst2}} \and  T. Goffrey\inst{\ref{inst1}}   \and C. Geroux\inst{\ref{inst1}}  \and M. Viallet\inst{\ref{inst3}}  \and D. Folini\inst{\ref{inst2}} \and  T. Constantino\inst{\ref{inst1}}  \and M. Popov\inst{\ref{inst2}}  \and R. Walder\inst{\ref{inst2}}  }
\institute{Astrophysics, College of Engineering, Mathematics and Physical Sciences, University of Exeter, EX4 4QL Exeter, United Kingdom \label{inst1} \and \'Ecole Normale Sup\'erieure de Lyon, CRAL (UMR CNRS 5574), Universit\'e de Lyon 1, 69007 Lyon, France\label{inst2} \and Max-Planck-Institut f\"ur Astrophysik, Karl Schwarzschild Strasse 1, 85741 Garching, Germany \label{inst3} } 

\titlerunning{Spherical-shell boundaries}
\authorrunning{J. Pratt et. al.}

\abstract
%\emph{Context:}
 {Studies of stellar convection typically use a spherical shell geometry.  The radial extent of the shell and the boundary conditions applied are based on the model of the star investigated.   We study the impact of different two-dimensional spherical shells on compressible convection.  Realistic profiles for density and temperature from an established one-dimensional stellar evolution code are used to produce a model of a large stellar convection zone representative of a young low-mass star, such as our sun at $10^6$ years of age. }
%\emph{Aims:} 
{We analyze how the radial extent of the spherical shell changes the convective dynamics that result in the deep interior of the young sun model, far from the surface.
In the near-surface layers, simple smaller-scale convection develops from the profiles of temperature and density.  A central radiative zone below the convection zone provides a lower boundary on the convection zone.  The inclusion of either of these physically distinct layers in the spherical shell can potentially affect the characteristics of deep convection. }
%\emph{Methods:}   
{We perform hydrodynamic implicit large-eddy simulations of compressible convection using the MUltidimensional Stellar Implicit Code (\music).
Because \music has been designed to use realistic stellar models produced from one-dimensional stellar evolution calculations, \music simulations are capable of seamlessly modeling a whole star.  
Simulations in two-dimensional spherical shells that have different radial extents are performed over tens or even hundreds of convective turnover times, permitting the collection of well-converged statistics.}
%\emph{Results:} 
{To measure the impact of the spherical shell geometry and our treatment of boundaries, we evaluate basic statistics of the convective turnover time, the convective velocity, and the overshooting layer.  These quantities are selected for their relevance to one-dimensional stellar evolution calculations, so that our results are focused toward studies exploiting the so-called 321D link.  We find that the inclusion in the spherical shell of the boundary between the radiative and convection zones decreases the amplitude of convective velocities in the convection zone.  The inclusion of near-surface layers in the spherical shell can increase the amplitude of convective velocities, although the radial structure of the velocity profile established by deep convection is unchanged.  The impact from including the near-surface layers depends on the speed and structure of small-scale convection in the near-surface layers.   Larger convective velocities in the convection zone result in a commensurate increase in the overshooting layer width and decrease in the convective turnover time.  These results provide support for non-local aspects of convection.}
{}

\maketitle

\section{Introduction}

Because convection underlies the fundamental processes of heat transport, mixing, shear, and the stellar dynamo, it also influences stellar evolution over long times.
Studies of stellar evolution typically use one-dimensional calculations that evolve physical quantities dependent on the radial position interior to a star.  These studies model the impact of convection on other physical quantities using one of the several variations on stellar mixing length theory, that depend on the local temperature gradient \citep[e.g.][]{vitense1953wasserstoffkonvektionszone, bohm1958wasserstoffkonvektionszone,abbett1997solar,trampedach2010convection,brandenburg2015stellar}.  Nevertheless, non-local processes can significantly influence stellar convection \citep[e.g.][]{spruit1997convection,gough1976calibration,canuto1997overshooting}.  Thus nonlinear hydrodynamic convection simulations that include a whole star are a key step to improving our understanding of stellar convection and eventually to improving models of stellar evolution.  

Whole-star simulations present considerable numerical challenges.  For that reason, numerical studies of hydrodynamic and magnetohydrodynamic convection are often performed in a simulation volume that is a spherical shell containing a portion of a star \citep[e.g.][]{gilman1983dynamically,cole2014azimuthal,yadav2013consistent,kapyla2011reynolds,kapyla2011effects,miesch2000three,grote2001dynamics,quataert2000convection,nelson2011buoyant,simitev2011turbulent}. 
 A spherical shell refers simply to any portion of a spherical domain that is limited both in angular and radial extent.  A spherical shell can be two-dimensional or three-dimensional, and sometimes also is called a spherical wedge.  
Several earlier works \citep{cole2016robust,mitra2009turbulent,heimpel2005numerical} have studied the impact of the spherical shell geometry and boundaries on the dynamo. The goal of this work is to produce a similar study of the impact of the spherical shell geometry on compressible convection without the fundamental influence of magnetic fields.

Although the impact of the spherical shell geometry on compressible
convection has not been studied, much fundamental work has been done to study the
 extent that convective plumes overshoot the boundary into the stable region.
 Earlier work on overshooting during compressible
convection has concentrated primarily on direct numerical simulation in a local
domain with Cartesian geometry in two
\citep{roxburgh1993numerical,hurlburt1994penetration} and three 
dimensions \citep{ziegler2003box,
brummell2002penetration,brummell2010dynamo}. Typically a
convectively unstable layer is stacked on top of a stable
layer.  The structure of this model is formed using a piecewise
continuous 2-layer polytropic stratification in the density,
temperature, and thermal diffusivity.
Direct numerical simulations using the anelastic approximation rather
than compressible convection have also been performed in a
two-dimensional cylindrical geometry \citep{rogers2005penetrative,
rogers2006numerical}; these works used a realistic stellar stratification rather than a polytropic stratification, but the thermal diffusivity was artificially increased.
Several other early studies have produced large eddy simulations, typically using a
standard Smagorinsky subgrid scale model, in a local domain with
Cartesian geometry
\citep{xie1993large,singh1995three, singh1998study,saikia2000examination,pal2007turbulent}. 
 Direct numerical simulation is a
sensible choice when the focus is dynamo action, because when a magnetic
field is present, small-scale motions feed back on large scale motions
in ways that are not completely understood \citep{miesch2015large}. 
However when hydrodynamic convection is the focus, the largest scales can potentially be the most important.

Because non-local effects of convection and coupling between physically different regions in a star can potentially play a significant role in the basic character of convective processes \citep[e.g.][]{brun2011modeling,latour1981stellar,spruit1997convection}, the necessary radial extent of the spherical shell is an open question.
 For practical reasons, simulations of stellar convection in a spherical shell often include a minimal radial extent of a modeled convection zone; neighboring, physically distinct zones of the star are neglected, and closed boundary conditions are applied at the top and bottom of this convectively unstable region of the star.
   In this work we compare results obtained when the interaction between a central radiative zone and an interior convection zone, and between an interior convection zone and near-surface layers is allowed.  We allow or disallow this coupling by either including these extra zones of the star in the spherical shell, or excluding them and applying a closed boundary condition.  Our motivation is to establish a physically reasonable position for the spherical shell and treatment of the boundaries that can serve as a basis for future studies of two- and three-dimensional convection at the bottom of the stellar convection zone, and overshooting of the boundary to the radiative zone.

\begin{figure*}[h]
\begin{center}
\resizebox{3.5in}{!}{\includegraphics{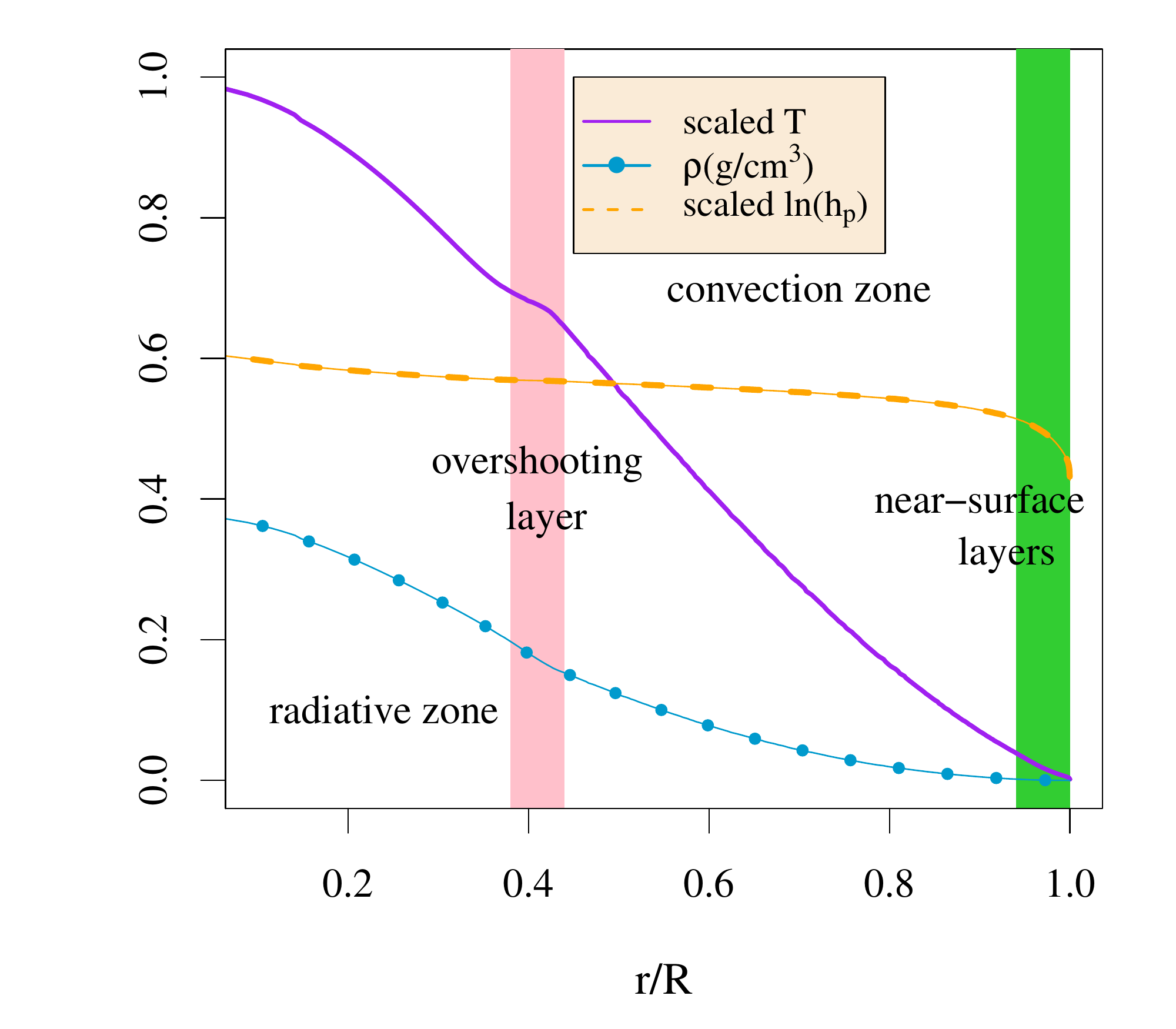}}
\caption{Diagram of the radial structure of our young sun model, showing the profiles of the natural log of the pressure scale height $h_p$, the temperature $T$, and density $\rho$.  The pressure scale height and temperature are scaled by arbitrary values in order to appear on the scale of this diagram.  $R$ is the total radius of the young sun.  Average radial widths shown for physical zones are predicted by the young sun model calculated with the Lyon stellar evolution code.
\label{figdiagram}}
\end{center}
\end{figure*}

In contrast to other solar and stellar simulations, the physical model of a star that we study in this work only allows for convection; the possibility of studying additional physical effects such as rotation, the effect of a tachocline, chemical mixing, or magnetic fields is intentionally omitted from the current study.  This simplifies the boundaries between the lower radiative zone, the convection zone, and the near-surface layers.   We study a prototypical model of a young, low-mass star: the young sun.  The stellar radius of our young sun model is approximately three times larger than the current sun, it is one solar mass, and has homogeneous chemical composition.  The radial profiles of density and temperature for the young sun model are typical for a pre-main sequence star that is no longer accreting and is gradually contracting.  The luminosity is increasing with the interior radius of the star.
Fig.~\ref{figdiagram} sketches the radial structure of the young sun model.    A central radiative zone below a large convection zone is expected based on the radial entropy profile and an evaluation of the Schwarzschild criterion.    An overshooting layer, where mixing of physically different flows from the radiative and convection zones can take place, can only be defined dynamically; this will be discussed in detail in Section~\ref{sectionovershoot}.
   Most importantly for this work, the young sun model has a large convection zone; it is convectively unstable over $1.2 \cdot 10^{11}$ cm of the total radius of $2.13 \cdot 10^{11}$ cm.  This large convective envelope allows us to study deep stellar convection, far from the physically complicated near-surface layers.  We term the type of convection analyzed in this work ``stellar convection'' because both the stratification in the density and the temperature gradient that drives the convection are non-uniform and not linearly dependent on the radius.  We define the near-surface layers as the portion of the star where the pressure scale height, $h_p=-p/(\partial p/\partial r)$, drops dramatically.   Further details of the young sun model will be discussed in Section 2. 

The most thoroughly studied type of convection is Rayleigh-B\'enard convection, which is driven by a uniform linear gradient of temperature, carefully controlled in a laboratory environment.  In Rayleigh-B\'enard convection, patterns can be identified \citep{bodenschatz2000recent,weiss2012pattern,emran2015large}.  In a star, the gradients of temperature and density are generally non-uniform and vary non-linearly over the full stellar radius. 
The radial variation of temperature and density for our young sun model is shown in Fig.~\ref{figdiagram}.   During convection driven by non-uniform gradients, although coherent structures such as plumes and convection rolls form, regular patterns cannot be identified.  Instead the diagnostics we use to evaluate the impact of the spherical shell geometry and boundaries are velocity statistics gathered over long periods of steady convection.

This work is structured as follows.  In Section 2 we discuss the simulation framework by outlining the physical and numerical models used by \music (MUltidimensional Stellar Implicit Code).  In Section 3 we discuss the details of several spherical shells that we formulate.  In Section 4 we compare simulations performed in different spherical shells with different radial extents on the basis of three different statistics related to the convective velocities.
In Section 5 we discuss the implications of these results for our planned future work on deep stellar convection and convective overshooting.

\section{Simulations}

Convection is typically modeled using one of three approaches:  the Boussinesq hydrodynamic equations, the anelastic hydrodynamic equations, or the compressible hydrodynamic equations \citep[for a concise summary see][]{glatzmaier2014introduction}.   While the Boussinesq and anelastic approximations are simpler and numerically more efficient,  there is reason to believe that these physical approximations could affect the realism of basic results in simulations of a whole star \citep{lecoanet2014conduction,verhoeven2015anelastic}.  In this work we use the compressible hydrodynamic equations.

The \music code solves the inviscid compressible hydrodynamic equations for density $\rho$, momentum $\rho \vec{u}$, and internal energy $\rho e$:
\begin{eqnarray} \label{densityeq}
\frac{\partial}{\partial t} \rho &=& -\nabla \cdot (\rho \vec{u})~,
\\ \label{momeq}
\frac{\partial}{\partial t} \rho \vec{u} &=& -\nabla \cdot (\rho \vec{u} \vec{u}) - \nabla p + \rho \vec{g} ~,
\\ \label{ieneq}
\frac{\partial}{\partial t} \rho e &=& -\nabla \cdot (\rho e\vec{u}) +p \nabla \cdot \vec{u} + \nabla \cdot (\chi \nabla T) .
\end{eqnarray}
Here the thermal conductivity $\chi=16 \sigma T^3/3 \kappa \rho$ is defined using  the Stefan-Boltzmann constant $\sigma$ and the Rosseland mean opacity  $\kappa$.
The thermal conductivity is thus realistic to the young sun model, and has not been enhanced \citep[for examples of ways that the thermal conductivity may be enhanced, see][]{browning2004simulations,rogers2006numerical,tian2009numerical,strugarek2011magnetic}.
A major feature of our simulations is the use of an equation of state and realistic opacities that are standardly used in one-dimensional stellar evolution calculations. Opacities are interpolated from the OPAL \citep{iglesias1996updated} and \citet{ferguson2005low} tables, which cover a range in temperature suitable for the description of the entire structure of a low-mass star. The compressible hydrodynamic equations \eqref{densityeq}-\eqref{ieneq} are closed by determining the gas pressure $p(\rho,e)$ and temperature $T(\rho,e)$ from a tabulated equation of state for a solar composition mixture.
This equation of state accounts for partial ionization of atomic species by solving the Saha equation, and neglects partial degeneracy of electrons; it is suitable for the description of our  solar model at a young age. The initial state for \music simulations is produced using data extracted from a one-dimensional model calculated from the Lyon stellar evolution code \citep{baraffe1991evolution, baraffe1997evolutionary,baraffe1998evolutionary}, which uses the same opacities and equation of state implemented in \music.
 In eq. \eqref{momeq}, $\vec{g}$ is the gravitational acceleration, a spherically-symmetric vector identical to that used in the Lyon stellar evolution code, and not evolved by our simulations.  Thus \music simulation results
should contribute to the 321D link \citep{david20143d, arnett2009turbulent}, \emph{i.e.} the effort to improve one-dimensional stellar evolution models by studying critical short phases using stellar hydrodynamics in two and three dimensions.

Early work to develop the \music code was reported-on in detail in \citet{viallet2011towards,viallet2013comparison}.  
Time integration is implicit in order to permit time steps larger than the Courant-Friedrichs-Lewy (CFL) limit for time-explicit methods.
  Convergence of an implicit scheme can be computationally demanding.  In the present study, the system of equations is discretized in time using a Crank-Nicolson scheme.
To integrate the compressible hydrodynamic equations in time, a Jacobian free Newton-Krylov (JFNK) solver \citep{knoll2004jacobian} is employed.  Instead of storing a Jacobian, a JFNK method uses matrix-vector products that can be estimated efficiently with finite difference methods.   In \music, a physics-based preconditioner \citep{park2009physics} targets the physical processes that cause the system to become stiff.  In practice, this preconditioning matrix is calculated as a semi-implicit solution of the system \citep{maximepaper}.  In all simulations in this work, we limit the time step by requiring a general CFL number be less than 10, while a CFL number based on simple advection is restricted to 0.5. This produces good convergence of relevant basic quantities, such as average kinetic energy. The development of efficient implicit solvers is the subject of on-going research \citep[e.g.][]{chacon2008optimal,main2014second,poedtsvolume2016}.
    
The spatial discretization of equations \eqref{densityeq}-\eqref{ieneq} is accomplished using a staggered grid and a finite volume approach.  Physical quantities are interpolated to the grid
using an upwind limited interpolation similar to the monotone upwind schemes for conservation laws (MUSCL) method \citep{thornber2008improved}. This method employs the well-known van Leer flux limiter \citep{roe1986characteristic}.
MUSIC is designed as a large-eddy simulation (LES).  In the present work no additional viscosity is applied, either through fixed coefficients or subgrid-scale modeling.   This is a common tactic for astrophysical hydrodynamics, in order to obtain the minimum possible dissipation \citep[for a similar discussion see][]{miesch2015large}.
 The \music code has been benchmarked \citep{tompaper} against standard hydrodynamic problems that isolate fundamental physical mechanisms relevant to stellar hydrodynamics, including an ideal Rayleigh-Taylor instability, an ideal Kelvin-Helmholtz instability, and the Taylor-Green vortex.  For these standard hydrodynamic problems \music produces the expected results, and exhibits the expected numerical convergence properties.
    
\section{Spherical Shell Simulation Volumes}
\subsection{Resolution}

The compressible hydrodynamic equations \eqref{densityeq}-\eqref{ieneq} are solved in a two-dimensional spherical shell using spherical coordinates:  radius $r$ and colatitude $\theta$.  In studying the impact of the spherical shell on compressible convection we examine only two-dimensional convection.  Two-dimensional convection is well-known to produce higher velocity structures than three-dimensional convection \citep[see for example][]{meakin2007turbulent}.  Boundary effects are also larger in two dimensions than in three.  Thus a study on the impact of boundary conditions and spherical shells on two-dimensional convection is the extreme case, and our results will also be relevant for three-dimensional convection simulations.

We consider the suite of two-dimensional simulations of spherical shell convection summarized in Table~\ref{simsuma}.  In this table, the radial position and extent of the spherical shell are shown in units of the total stellar radius $R$.     Most of the simulations use a uniform grid characterized by a fixed radial spacing $\Delta r$.   The low resolution simulations (Low1-8) use radial grid spacing $\Delta r /R \approx 2.8 \cdot 10^{-3}$.  The high resolution simulations (Hi1-4) use radial grid spacing  $\Delta r/R \approx 1.4 \cdot 10^{-3}$.  The extra-high resolution simulation ExH2 uses radial grid spacing  $\Delta r/R = 7 \cdot 10^{-4}$.   Each of these simulations that uses a uniform grid also uses a constant energy flux across the upper surface.

To explore the impact of certain boundary conditions on the energy, described in detail in Section \ref{secbc}, a higher resolution of the near-surface layers is required.
To accomplish this, the remaining two simulations, Low9, and Hi5, use a spliced grid.  Our spliced grid is composed of a uniform grid with fixed grid spacing $\Delta r$ for $r/R \leq 0.94$.  In the near-surface layers, for $r/R > 0.94$, a non-uniform grid that is decreasing radially toward the surface is spliced on top of the uniform grid.  The non-uniform portion of the grid is defined by a geometric sequence $\Delta r_i = \Delta r_{i+1}/1.05$ \citep{chrispaper}.  The grid spacing in the uniform portion of the grid in simulations Low9, and Hi5 is identical to that in simulations Low1-8 and Hi1-4, respectively.
  The spliced grid of simulation Hi5 allows for a resolution of $\Delta r/h_p(r) \approx  0.36$ at the surface; for comparison, the uniform grid in simulation Hi4, has $\Delta r/h_p(r) \approx 0.67$ at the surface.  For either the spliced grid or the uniform grid, our grid-spacing in the near-surface layers is too low to resolve the complex physics in these layers with any precision.  Our motivation for modeling these layers is solely to provide a simple physically-motivated open boundary condition \citep[e.g.][]{kapyla2010open,cameron2011decay} on the interior convection zone.  Such an open boundary condition allows the exchange of momentum, density, and thermal fluctuations with other zones of the star; open boundary conditions are highly desirable when dealing with large-scale flows.

\begin{figure*}[h]
\begin{center}
\resizebox{3.5in}{!}{\includegraphics{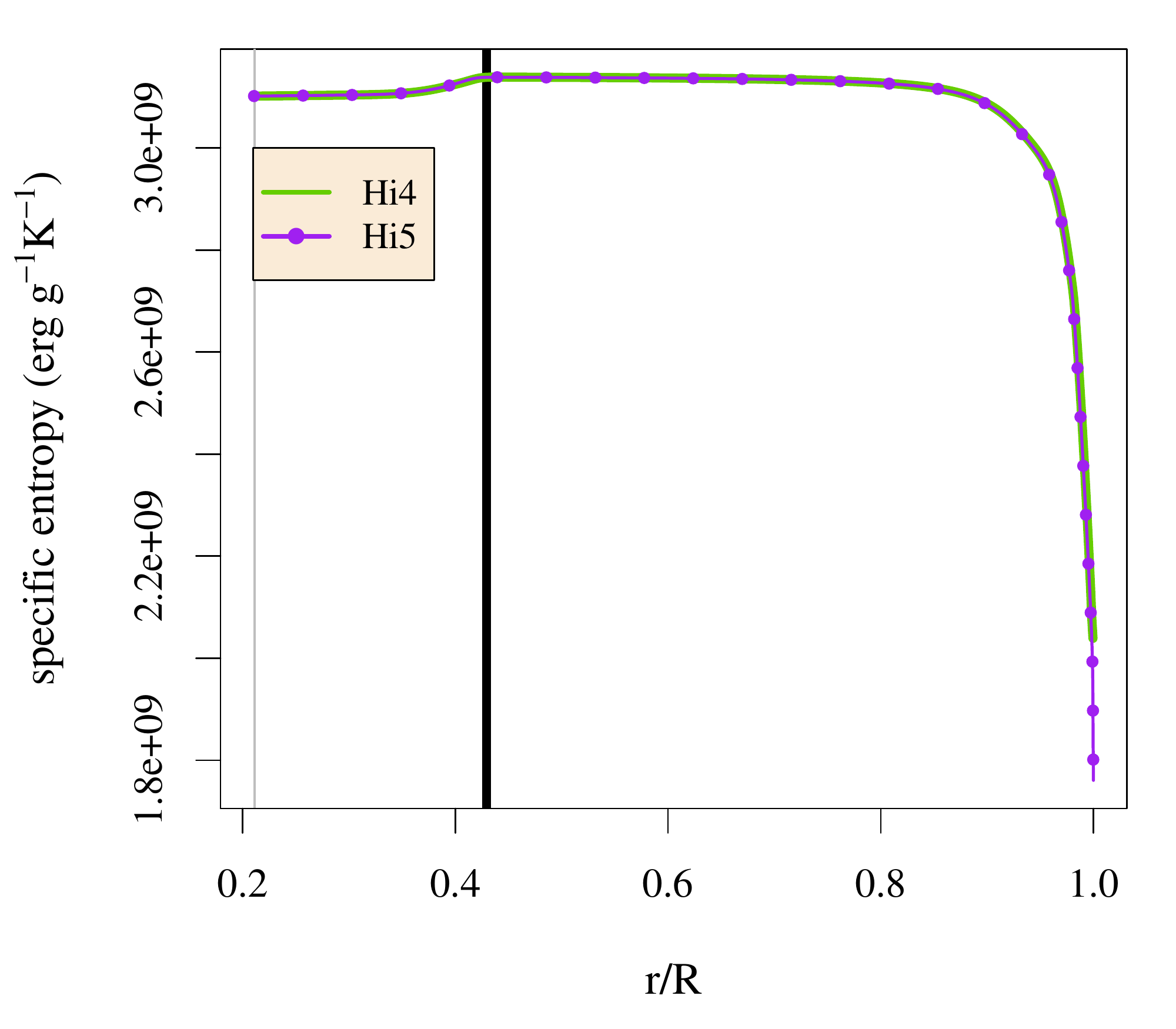}}
\caption{Time-averaged radial profile, produced from a volume-weighted average in the angular direction, of specific entropy in simulations Hi4 and Hi5.  A heavy vertical black line marks the boundary between the radiative and convection zones, determined from the radial profile of entropy and the Schwarzschild criterion produced by the one-dimensional stellar model. A grey vertical line marks the inner radial boundary of the spherical shell at $0.21$ of the stellar radius.
 \label{vrmssurf_entropy}}
\end{center}
\end{figure*}

  The resolutions we employ are comparable to earlier LES studies \citep[e.g.][]{brun2009numerical};
   we note that simulation Hi2 has a total grid size of $608 \times 512$.  
 By not pursuing extremely high resolution simulations for this benchmarking study of large-scale convection in a spherical-shell geometry, we are able to simulate over long times.  
 Because the one-dimensional stellar evolution model of the young sun stops at the photosphere, defined by an optical depth $\tau=2/3$, our simulations do not include the usual increase in entropy in the stable atmospheric layers \citep[e.g.][]{magic2016,trampedach2014improvements,abbett1997solar}.  The preceding drop in entropy in the near-surface layers is under-resolved.   However, the entropy jump at the bottom of the convection zone is resolved in our simulations. The entropy profile for simulations Hi4 and Hi5 is shown in Fig. \ref{vrmssurf_entropy}.   
 Because the characteristic length scales, velocities, and thermal diffusivity vary throughout the radius of a star, the Rayleigh number and Reynolds number are not specified in a general sense for these simulations.  We note that such nondimensional parameters can potentially affect convective heat transport \citep{ahlers2001prandtl,kerr2000prandtl} and convection dynamo action \citep{simitev2005prandtl}.  In \music, numerical truncation errors contribute to both the viscosity and thermal diffusivity.  In addition \music simulations include an explicit thermal diffusivity related to the thermal conductivity in eq. \eqref{ieneq}, so that the Prandtl number is everywhere less than one.  The simulations possible with a code like \music are still many orders of magnitude away in parameter space from
the highly turbulent conditions likely to be found in realistic stellar convection zones. The LES results should therefore be viewed merely as indicators of the properties of realistic stellar flows.  

\subsection{Boundary conditions \label{secbc}}

Because the placement of boundaries and the choice of boundary conditions affect the physical outcome of
a hydrodynamic simulation, we use boundary conditions that are targeted to maintaining the original radial profiles of density and temperature. 
Each simulation volume begins at $20 \degree$ from the north pole, and ends $20 \degree$ before the south pole.  We impose periodicity on all physical quantities at the boundaries in $\theta$.  Radial boundaries require more sophisticated treatment.  In velocity, we impose non-penetrative and stress-free boundary conditions on the radial boundaries.

When the near-surface layers are included in the spherical shell and resolved using a spliced grid, in simulations Low9 and Hi5, a black-body radiation law is used on the outer radial boundary.  To radiate as a black-body, the energy flux is allowed to vary as $\sigma T_{\mathsf{s}}^4$ where $\sigma$ is the Stefan-Boltzmann constant and $T_{\mathsf{s}}(\theta,t)$ is the temperature along the surface.  This physically realistic boundary condition can only be effectively used when the steep temperature gradient near the surface is sufficiently resolved; otherwise it results in artificially high cooling rates.  In this work when a uniform grid is used, the energy flux and surface luminosity are held constant at the outer radial boundary at the correct value of the energy flux at that radius in the one-dimensional stellar evolution calculation; the surface luminosity of our model for the young sun is 2.32 times the luminosity of our current sun.  On the inner radial boundary the energy flux is always held constant at the value indicated by the one-dimensional stellar evolution data.  

To treat the density, we use the hydrostatic equilibrium boundary condition described by \citet{hsegrimm2015realistic}.  Integration of the equation for hydrostatic equilibrium produces two possibilities for the density stratification:
\begin{eqnarray}\label{eqhse1}
\rho_j (\theta) &=& \rho_0 (\theta) \exp{\left( -(r_j - r_0)~g_0 / (d p/d\rho)  \right)} ~,
\\
\rho_j (\theta) &=& \rho_0 (\theta) \exp{\left( - (r_j - r_0) / h_p \right)} ~.  \label{eqhse2}
\end{eqnarray}
Here $j$ is a grid index in the boundary cells, with $j=0$ indicating the last radial cell inside the spherical shell. Eq.~\eqref{eqhse1} assumes constant internal energy and constant radial acceleration due to gravity in the boundary cells.  Eq.~\eqref{eqhse2} assumes a constant pressure scale height in the boundary cells.  This type of boundary condition assumes an ideal gas at the boundaries.

When the boundary condition does not closely match the density stratification of the stellar model, an artificial boundary layer develops near the simulation boundary.  When the inner radial boundary of the spherical shell is in the radiative zone, we impose a constant radial derivative on the density at this boundary.  A constant derivative is a comparatively simple boundary condition, but for the young sun we find that it more accurately and smoothly maintains the density stratification than using the hydrostatic equilibrium boundary condition on the inner radial boundary.  When the inner radial boundary of the spherical shell is in the convection zone, we impose the hydrostatic equilibrium boundary condition on the density using eq.~\eqref{eqhse2}.  
  We find that in the lower convection zone, this hydrostatic equilibrium boundary condition minimizes the size of the boundary layer that would otherwise develop.

At the outer radial boundary we find that the hydrostatic equilibrium boundary conditions produce results indistinguishable from simply assuming a constant radial derivative for the density.  We apply a boundary condition on the density that maintains hydrostatic equilibrium using eq.~\eqref{eqhse1}.   The accuracy of a particular boundary condition depends on the placement of the boundary and the physics of the fluid flow near the boundary, which are dictated by the spherical shell geometry and the type of star.

\begin{table*}
\begin{center}
\caption{Parameters for two-dimensional hydrodynamic simulations of the young sun.
 \label{simsuma}
 }
\begin{tabular}{lccccccccccccccccccccccccccc}
\hline\hline
                      & inner radius ($R$)  & outer radius ($R$)  & $r_{\mathsf{o,top}}$ ($R$) & $w_{\mathsf{o}}$ (\% $R$)  & $\tau_{\mathsf{conv}}$($10^6$s) & time span ($\tau_{\mathsf{conv}}$) & grid type 

\\ \hline
Low1                  & 0.41 ({small})          & 0.91    & 0.50  & 9.4 & $ 3.21 \pm 2.10$ & 358  & uniform %& 176 %again9
\\ \hline
Low2                  & 0.31 ({small})          & 0.91     & 0.44 & 3.7  & $ 1.52 \pm 0.47$ & 465 & uniform  %& 220 %again8
\\ \hline
Low3                  & 0.21 ({medium})          & 0.94    & 0.45 & 4.0  & $ 1.40 \pm 0.34 $  & 686  & uniform %& 256 %pink/ill1
\\ \hline
Low4                  & 0.10  ({large})          & 0.97    & 0.45  & 2.9  & $ 2.66 \pm 0.75 $ & 338 & uniform %& 304 %purple/ill2
\\ \hline
Low5                  & 0.21 ({medium})           & 0.97      & 0.45 & 3.4  & $ 2.07 \pm 0.32$ & 330  & uniform %& 268 %chartreuse3/ill3
\\ \hline
Low6                  & 0.10 ({large})          & 0.94      & 0.45 & 4.2  & $ 1.61 \pm 0.66$ & 473  & uniform %& 296 %blue/ ill4
\\ \hline
Low7                  & 0.47  ({none})          & 1.00       & - & -  & $ 1.29 \pm 0.40$ & 432  & uniform  % & 236 %altsurf
\\ \hline
Low8                & 0.21 ({medium})          & 1.00      & 0.44 & 2.6 & $ 1.88 \pm 0.16$  & 68  & uniform % & 276 %again10
\\ \hline
Low9                  & 0.21 ({medium})         & 1.00     & 0.47 & 7.7  & $ 0.88 \pm 0.20 $& 69  & spliced % & 320 % chrisdata
\\ \hline  \hline
Hi1                   & 0.24  ({medium})         & 0.94   &  0.52 & 13.5  & $ 1.17 \pm 0.12 $  &  157  & uniform  % & 512 % chat/pink/ke1 **
\\ \hline 
Hi2                   & 0.10 ({large})          & 0.97     & 0.53 & 16.6  & $ 1.20 \pm 0.15 $  & 85  & uniform % & 608 % bade/purple/ke4 **
\\ \hline 
Hi3                   & 0.21  ({medium})         & 0.97     & 0.51 & 11.3  & $ 1.08 \pm 0.22$  & 163  & uniform  % & 536 % genu/char/ke2 **
\\ \hline
Hi4                  & 0.21  ({medium})         & 1.00     & 0.58  &  14.7 & $ 0.91 \pm 0.23$  &  49   & uniform % & 704 %grind/
\\ \hline  
Hi5                   & 0.21 ({medium})          & 1.00     & 0.54 & 13.1  & $ 0.72 \pm 0.07$ &  59  & spliced  % & 600 % redo/
\\ \hline  \hline
ExH2                & 0.10 ({large})           & 0.97    & 0.56 & 15.1 & $ 0.91 \pm 0.07 $  &  18  & uniform % & 1216 %dust/
\\ \hline\hline
\end{tabular}
\tablefoot{The inner and outer radius of the spherical shell are given in units of the total radius of the young sun $R$.  The amount of the radiative zone included is categorized in terms of a small, medium, or large amount, or none.  Diagnostic results are summarized, including the outer radius $r_{\mathsf{o,top}}$ of the overshooting layer, the width of the overshooting layer $w_{\mathsf{o}}$, and the convective turnover time $\tau_{\mathsf{conv}}$. The total time span and the type of grid used for each simulation is also indicated.  The type of grid is linked to the type of boundary condition at the surface.}
\end{center}
\end{table*}

\section{Results}

\subsection{Convective turnover time \label{secturntime}}
The convective turnover time is a fundamental parameter produced in one-dimensional stellar evolution studies and used in further modeling \citep[e.g.][]{kim1996theoretical,matt2015mass}.
It can be defined in a variety of nearly equivalent ways \citep[e.g.][]{landin2010theoretical,meakin2007turbulent}.
In this work we define a local convective turnover time at any position in our convection zone as 
\begin{eqnarray}\label{eqtaulocal}
\tau_{\mathsf{loc}} (r,\theta,t) = h_p(r,\theta,t)/ |v(r,\theta,t)|~.
\end{eqnarray}
  In this expression $h_p$ is the pressure scale height as a function of position and time; $|v|$ is simply the velocity magnitude.  Our stellar hydrodynamics model does not include rotation or mean-flows, and no mean-flows result in our simulations; the velocity magnitude in the convection zone is a purely convective velocity.
  A global time-scale $\tau_{\mathrm{global}}$ can thus be defined from the local convective turnover time by averaging over the entire convection zone:
\begin{eqnarray}\label{eqtauconv}
\tau_{\mathrm{global}}(t) =  \int_{r_{\mathsf{min}}}^{r_{\mathsf{max}}} \int_{\theta_2}^{\theta_1}  d \mathsf{V(r,\theta)}~ \tau_{\mathsf{loc}} (r,\theta,t)  ~/~ \int_{r_{\mathsf{min}}}^{r_{\mathsf{max}}} \int_{\theta_2}^{\theta_1}  d \mathsf{V(r,\theta)}
\end{eqnarray}
Here the integration is volume-weighted and $\mathsf{V(r,\theta)}$ is a volume element.  The integration covers the convection zone, between  $r_{\mathsf{min}}/R=0.47$ and $r_{\mathsf{max}}/R=0.85$, and the full angular extent of our spherical shells.  This lower limit is chosen as the deepest point within the convection zone of the young sun model that appears to be unaffected by the changing dynamics near the boundary to the radiative zone.  The upper limit is chosen as a point in the upper convection zone that is contained in all of our spherical shells; in the upper convection zone the contribution to the convective turnover time becomes small.  In practice the variation of the pressure scale height in the angular direction and in time is small; an expression based on an average of the pressure scale height divided by the root-mean-square of the radial velocity yields similar results to our definition in eq.~\eqref{eqtauconv}.  However capturing small variations is an objective of our simulations and this motivates the form of eq.~\eqref{eqtauconv}.  Ultimately the convective turnover time is time-averaged over the full time span, noted in Table~\ref{simsuma}, of the simulation to produce a single number representative of convection in the star, \emph{i.e.} $\tau_{\mathrm{conv}} = \langle \tau_{\mathrm{global}} \rangle_t$, where the brackets $\langle ...\rangle_t $ indicate a time average.

\begin{figure*}[h]
\begin{center}
\resizebox{3.5in}{!}{\includegraphics{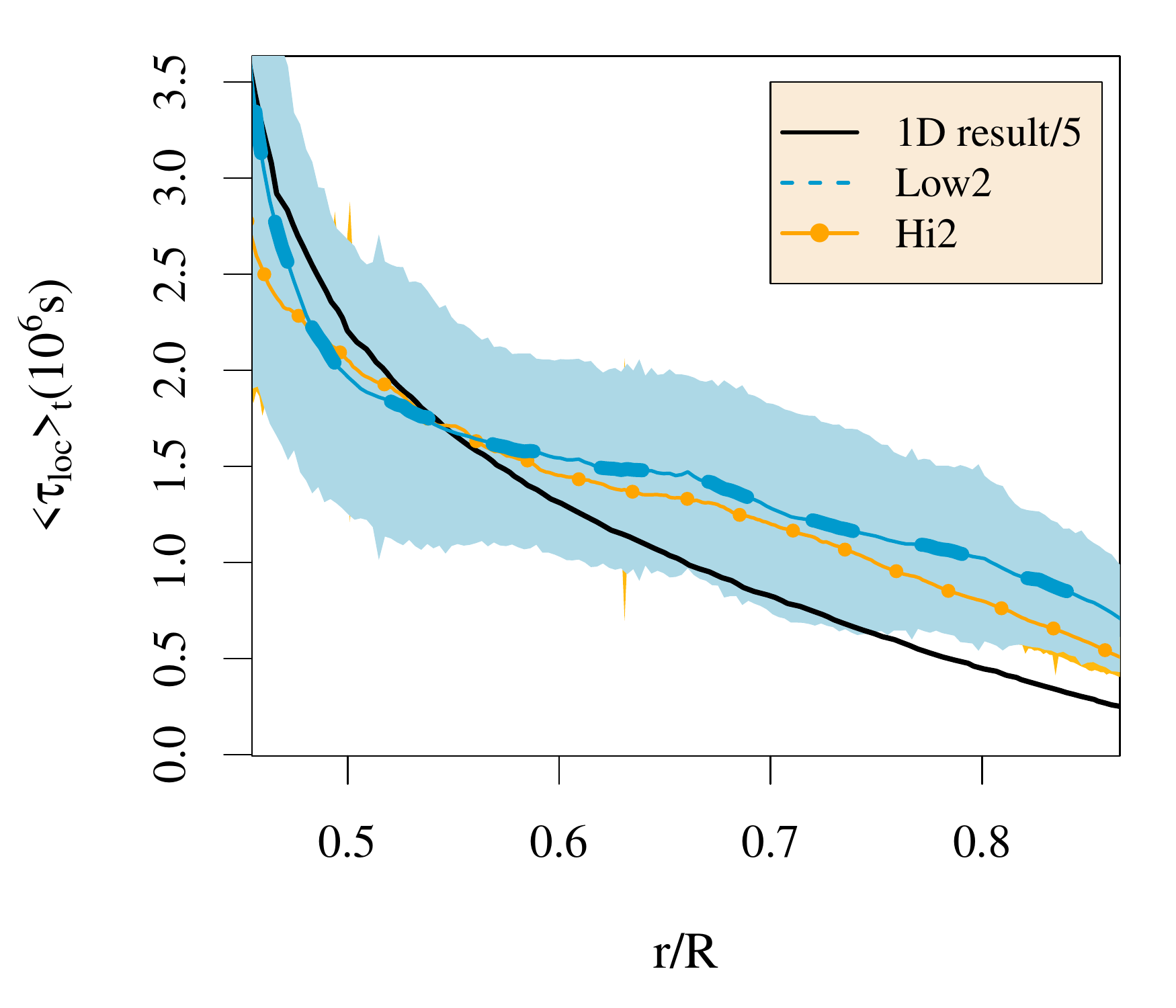}}
\caption{Radial profiles of $\langle \tau_{\mathsf{loc}} \rangle_t$ the local convective turnover time, averaged over the full simulation time in simulations Low2 and Hi2, compared with the value obtained from the one-dimensional Lyon stellar evolution code, after dividing by a factor of 5.  The shaded areas indicate one standard deviation above and below each of these time-averaged lines; the shaded areas are nearly identical for these two simulations.
\label{figpdftau}}
\end{center}
\end{figure*}

The radial profiles of $\langle\tau_{\mathsf{loc}}\rangle_t $ from simulations Low2 and Hi2, calculated using a volume-weighted average in $\theta$, are shown in Fig.~\ref{figpdftau}, and compared with a result from the one-dimensional Lyon stellar evolution code.  The result from the one-dimensional calculation uses a velocity produced by mixing length theory: $\langle\tau_{\mathsf{loc}}\rangle_t |_{\mathsf{1D}}\equiv h_p/v_{\mathsf{MLT}}$.   Two-dimensional hydrodynamic calculations produce a radial profile with a shape similar to the one-dimensional stellar evolution calculation.  However at the bottom of the convection zone $\langle\tau_{\mathsf{loc}}\rangle_t $ initially decreases more sharply than $\langle\tau_{\mathsf{loc}}\rangle_t |_{\mathsf{1D}}$.  In the upper convection zone, the two-dimensional hydrodynamic simulations display a radial profile of $\langle\tau_{\mathsf{loc}}\rangle_t $ that is flatter, and has a higher value.  The magnitude of $\langle\tau_{\mathsf{loc}}\rangle_t $ in the hydrodynamic calculations is smaller by approximately a factor of 5 than $\langle\tau_{\mathsf{loc}}\rangle_t |_{\mathsf{1D}}$. Because two-dimensional simulations produce higher velocities than three-dimensional simulations, we expect that the convective turnover time from three-dimensional simulations will lie between the one- and two-dimensional values.

Table~\ref{simsuma} summarizes $\tau_{\mathrm{conv}} = \langle \tau_{\mathrm{global}} \rangle_t$, as well as the standard deviation of $\tau_{\mathrm{global}}(t)$, statistics calculated from data sampled at a fixed time-interval throughout our statistically steady convection simulations.  The standard deviation is used in this context as an error estimate on the mean.  In this table the total simulation time, i.e. the span of time in steady-state convection for which each simulation has been followed, is also indicated in units of the convective turnover time.  The values of $\tau_{\mathrm{conv}}$ for simulations Low1-9 are mildly larger than for simulations Hi1-5.  This reflects that lower velocities result from lower resolution simulations where numerical dissipation is larger. For both sets of simulations Low1-9 and Hi1-5, $\tau_{\mathrm{conv}}$ is on the order of $10^6$s.  The one-dimensional Lyon stellar evolution code produces a larger convective turnover time on the order of $5 \cdot 10^6$s.

\begin{figure*}[h]
\begin{center}
\resizebox{3.5in}{!}{\includegraphics{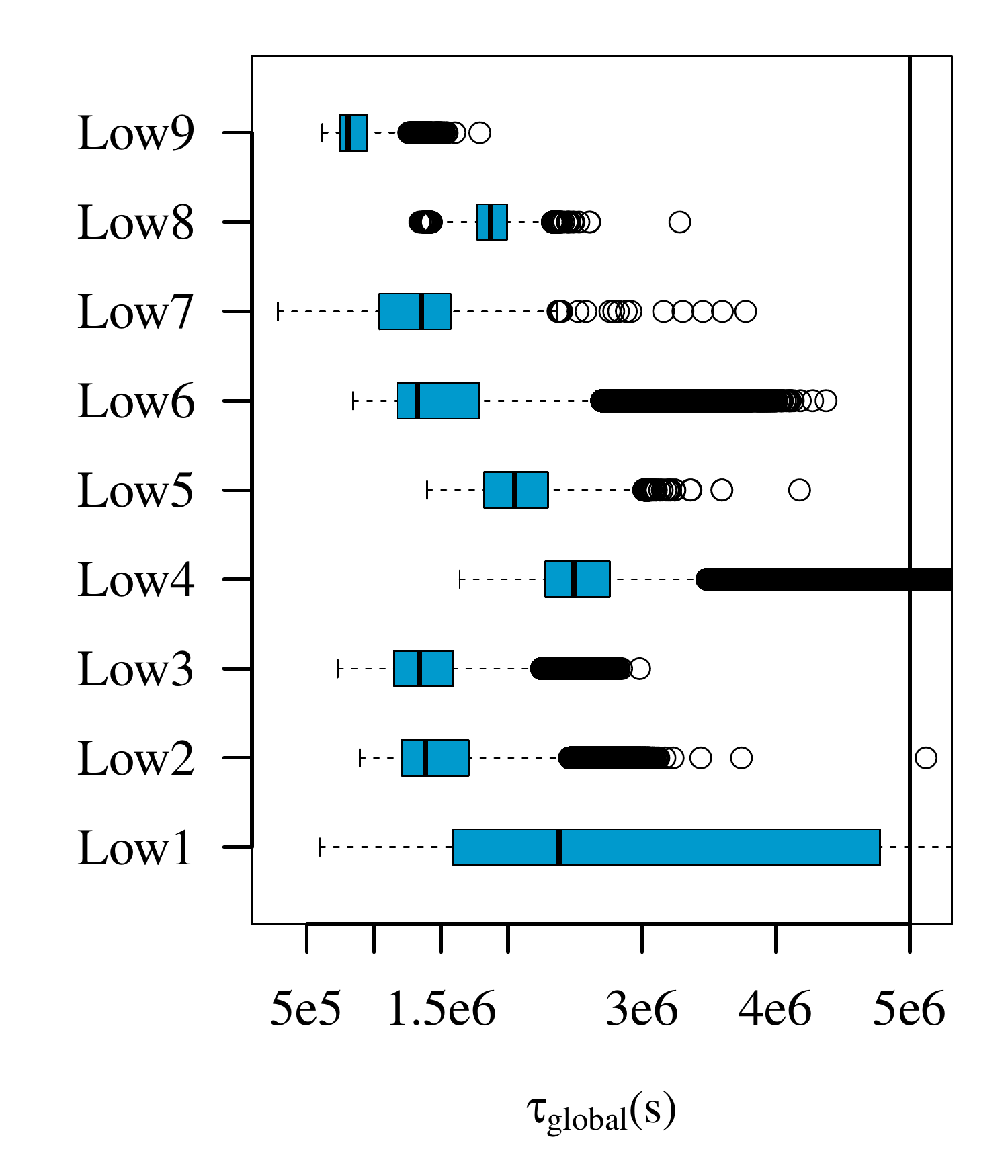}}\resizebox{3.5in}{!}{\includegraphics{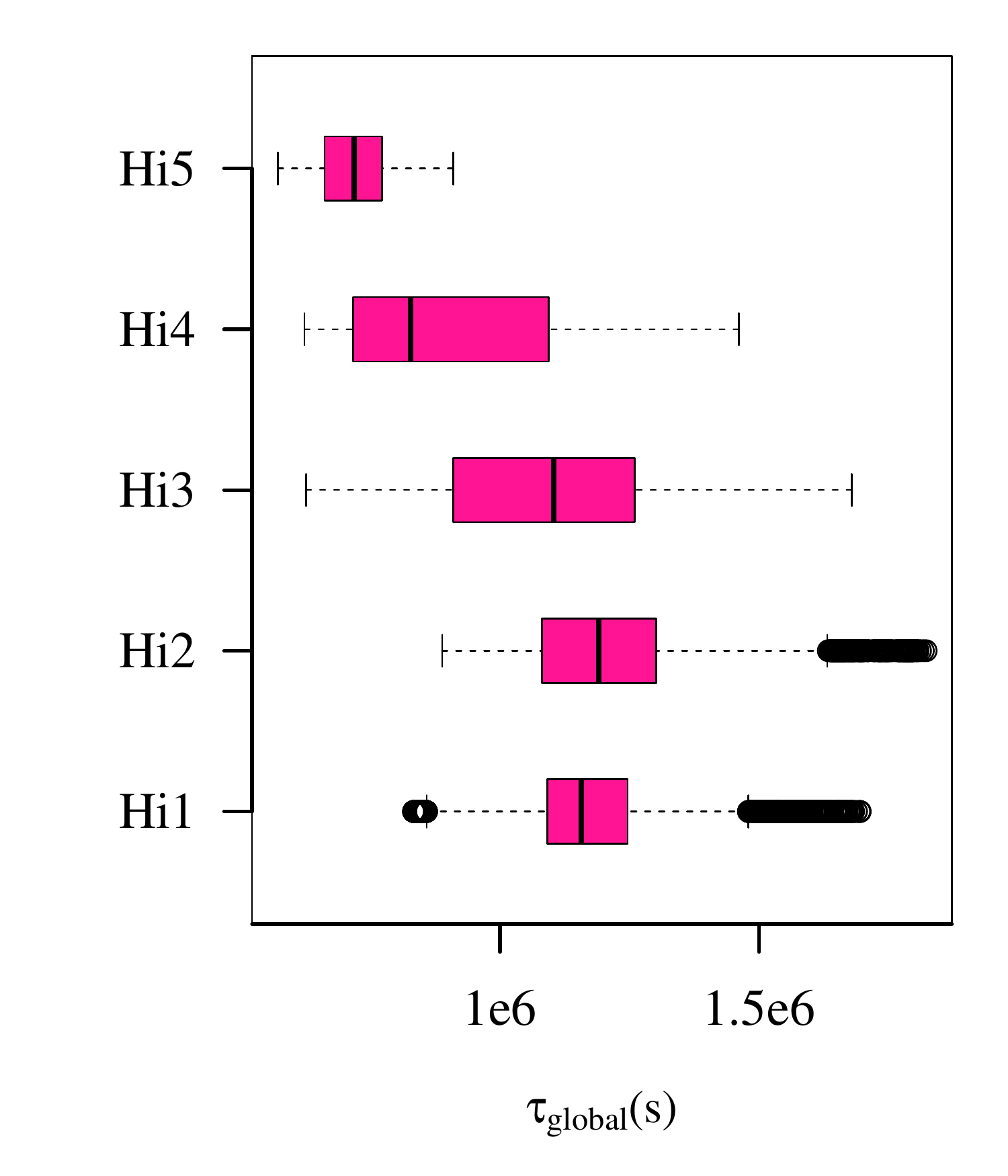}}
\caption{Standard Tukey box plot of the global time-scale $\tau_{\mathrm{global}}$ sampled at a fixed time-interval throughout the simulation (left) Low1-9 and (right) Hi1-5. A box plot \citep[e.g.][]{spitzer2014boxplotr,mcgill1978variations} is designed to show characteristics of the spread of the data. 
 The line in the middle of the box is the median. The box itself represents the middle 50\% of the data, so that the edges are the 25th and 75th percentiles.  The whiskers, i.e. error bars, represent the extent of the data when outliers are discounted.  Outliers are represented by circles. For a Tukey boxplot, outliers are defined as data that lies further than 1.5 times the interquartile range from the box.
 The vertical black line at $5 \cdot 10^6$s marks the value of $\tau_{\mathrm{conv}}$ that we obtain from mixing length theory.
\label{figboxplottau}}
\end{center}
\end{figure*}

To understand this difference between the time-averaged values of $\tau_{\mathrm{conv}}$, it is useful to analyze the spread of $\tau_{\mathrm{global}}$ sampled at a fixed time-interval on the order of $ \tau_{\mathrm{conv}}/10^{3}$ during our simulations.
 Fig.~\ref{figboxplottau} shows the full spread of this data as a box plot.  In the left panel of this figure, simulations Low1-9 are shown on the vertical axis, while the spread of $\tau_{\mathrm{global}}$ data is shown on the horizontal axis.  The high value produced by the one-dimensional Lyon stellar evolution code is only occasionally reached.   In the right panel of this figure, similar box plots are shown for higher resolution simulations Hi1-5.  
 By comparing simulations with identical grid spacing, we eliminated the effect of resolution.  We are then able to observe that, although different spherical shells of the star are covered, 
   the data sets of $\tau_{\mathrm{global}}$ strongly overlap. We also observe that simulation Low1, which includes the boundary at the bottom of the convection zone, but does not include the radiative zone, experiences a larger variation in time for $\tau_{\mathrm{global}}$ than do simulations that include the radiative zone.

\subsection{Radial structure and amplitude of convective velocities in the young sun}

\begin{figure*}[h]
\begin{center}
\resizebox{2.2in}{!}{\includegraphics{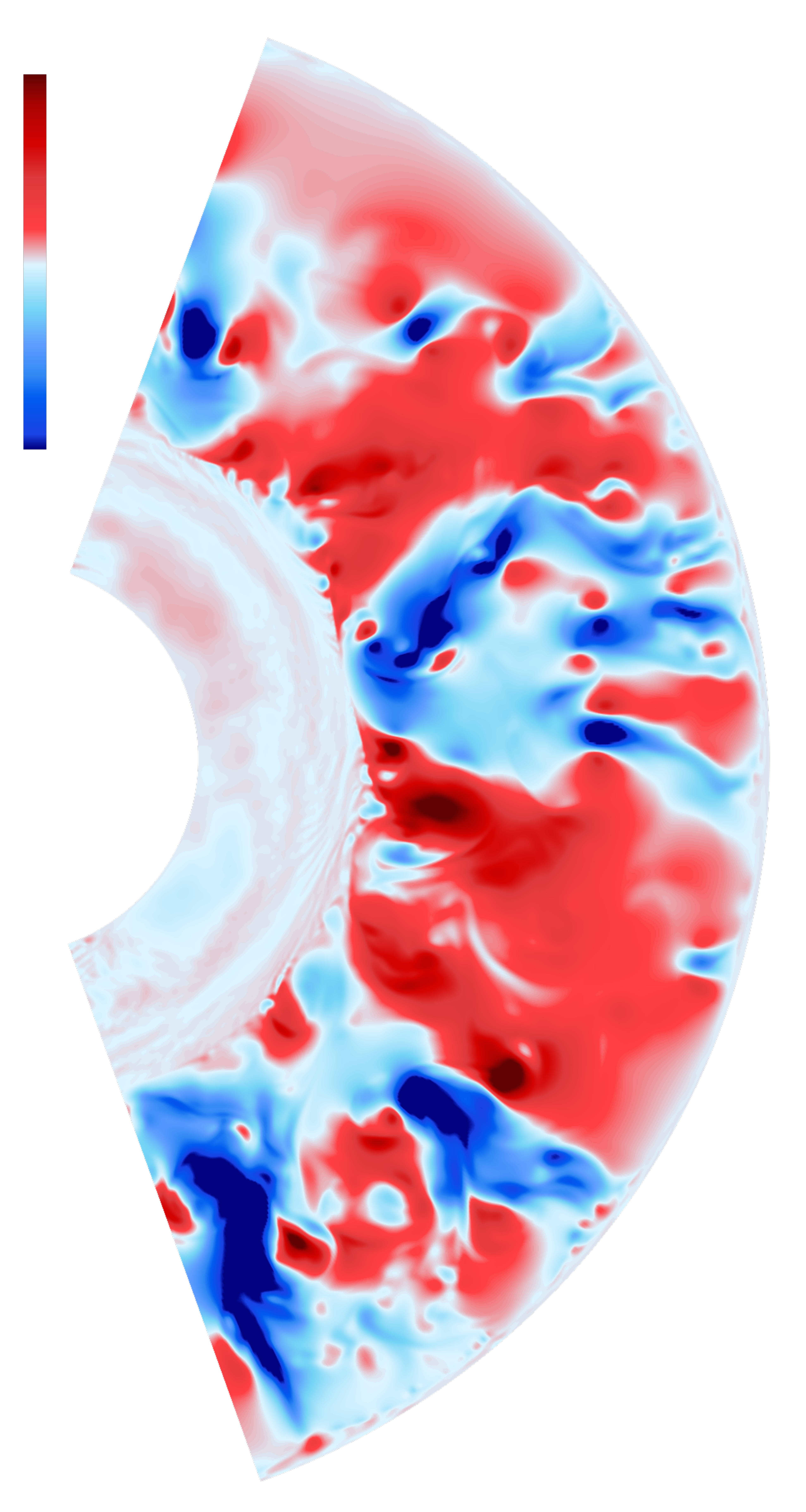}}\resizebox{2.35in}{!}{\includegraphics{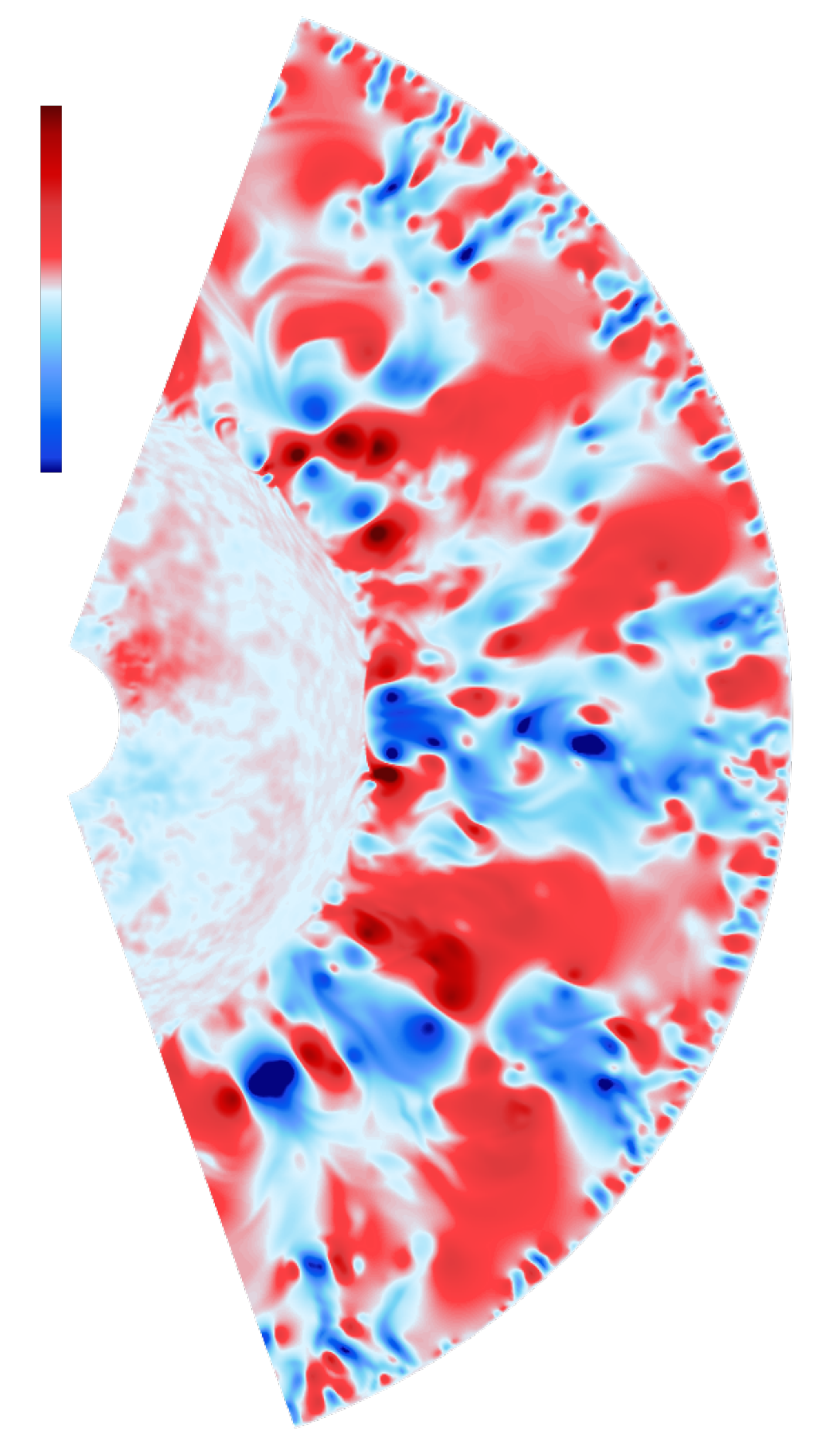}}
\caption{Instantaneous radial velocity in simulations of the young sun. The left panel shows a typical snapshot from simulation Hi1, which excludes the near-surface layers and the lower radiative zone.  The right panel is typical of simulation Hi2, which includes a large extent of the radiative zone.   Smaller-scale convective motions are clearly visible in the near-surface layers of simulation Hi2.  These simulations have identical temperature and density profiles, grid spacing, and boundary conditions.  They are both visualized during steady convection. The sole difference is the position of the simulation boundaries in the radial direction.  The color schemes are identical: blue indicates an inward flow, while red indicates an outward flow.
\label{velmagviz}}
\end{center}
\end{figure*}

In our model of the young sun, steady convection is characterized by wide regions of upflow velocity, and thinner regions with faster downflows.  In contrast to three-dimensional convection, in two-dimensions the convection rolls are all aligned.   Fig.~\ref{velmagviz} illustrates this structure in the radial velocity for simulations Hi1 and Hi2.  Simulation Hi2 includes both more of the near-surface layers and more of the radiative zone than simulation Hi1.
From this figure it is clear that Hi2 has developed small-scale convection in the near-surface layers that is absent in simulation Hi1.  The additional extent of the radiative zone permits the growth of more waves in the radiative zone of simulation Hi2.  However, the structure of flows in the bulk of the convection zone are not visibly distinguishable between simulations Hi1 and Hi2.   

Radial positions of convection rolls are time dependent.  Because of this it is meaningful to examine radial profiles of the time average of the root-mean-square (RMS) velocity $\langle v_{\mathsf{RMS}}\rangle_t$ which can be conveniently linked with helioseismology data and mixing-length theory predictions \citep[e.g.][]{miesch2012amplitude}.
These radial profiles of velocity can provide detail about the dynamic radial structure of the young sun.
   
\subsubsection{Coupling to a central radiative zone: impact on velocity}

\begin{figure*}[h]
\begin{center}
\resizebox{3.5in}{!}{\includegraphics{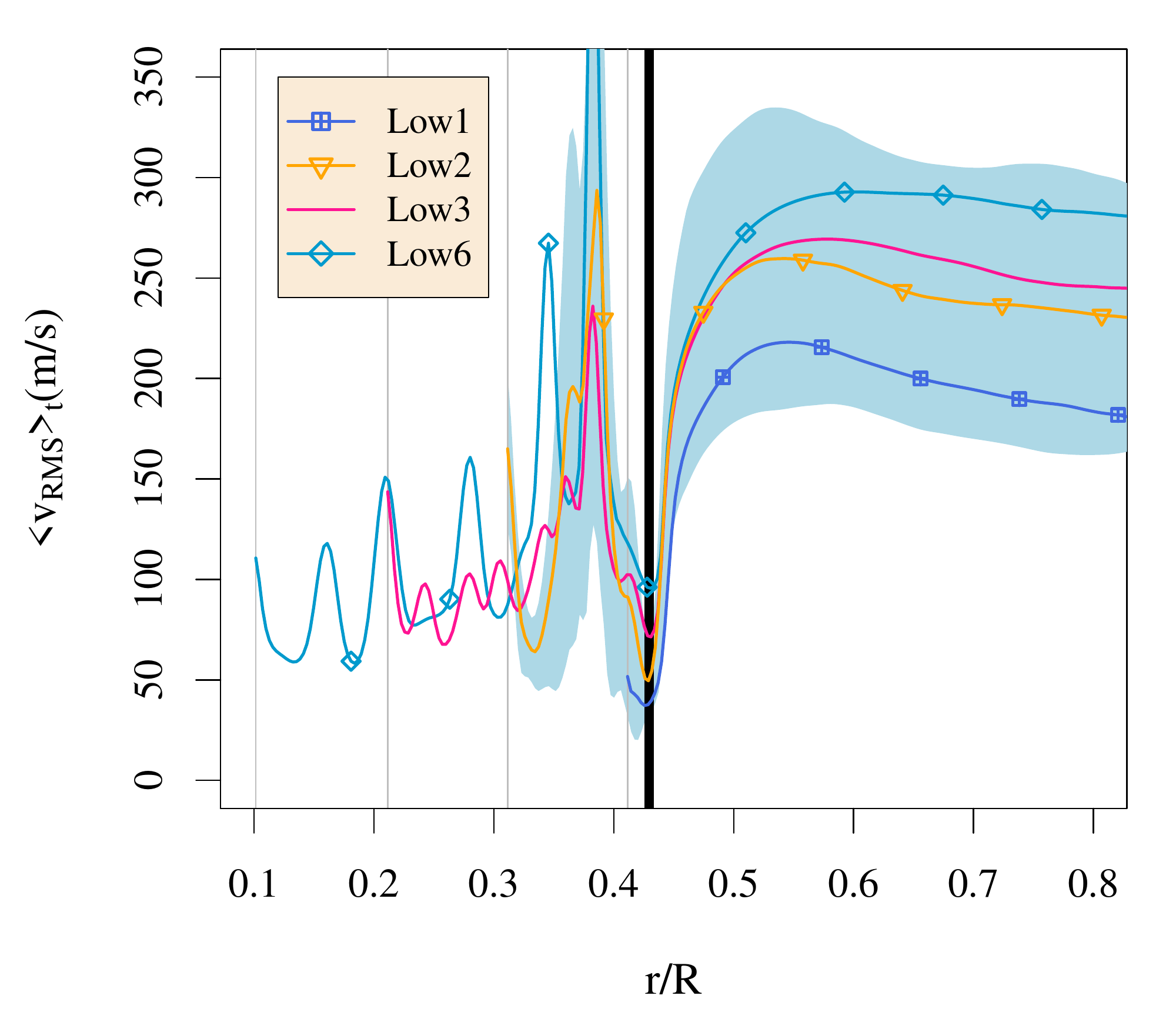}}\resizebox{3.5in}{!}{\includegraphics{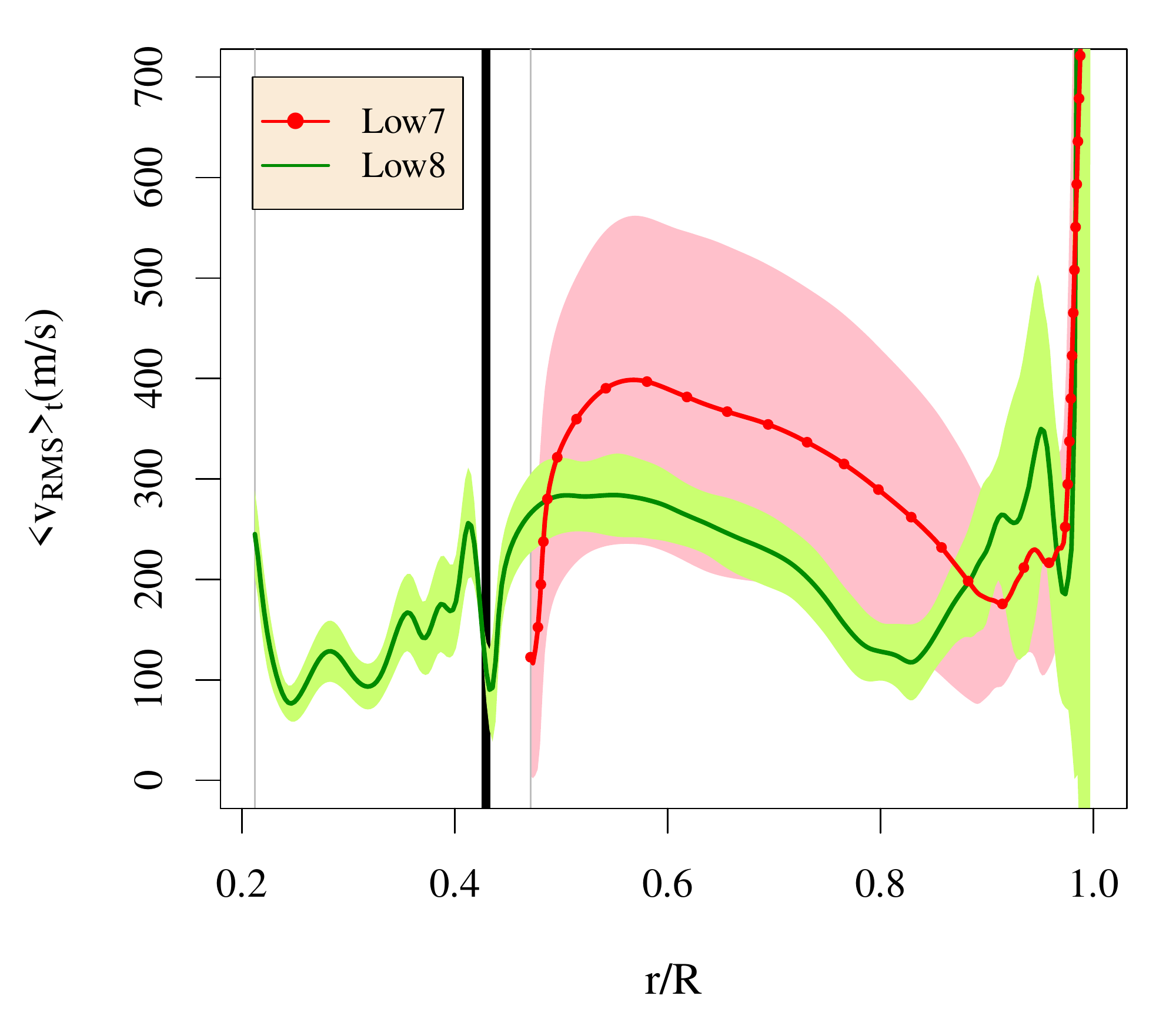}}
\caption{(Left) time-averaged radial profile of RMS velocity in four low-resolution simulations that exclude the near-surface layers but include a different extent of the radiative zone.
(Right) time-averaged radial profile of RMS velocity in simulations Low7 and Low8, that include the near-surface layers up to the surface at $r/R=1.0$.
   The shaded areas indicate one standard deviation above and below the time-averaged lines; in the left panel the shaded area for simulation Low2 is shown.  A heavy vertical black line marks the boundary between the radiative and convection zones, calculated from the radial profile of entropy produced by the one-dimensional stellar evolution calculation and the Schwarzschild criterion.  Grey vertical lines mark the inner radial boundary of the spherical shell simulation volume for each simulation shown.
 \label{vrmsconv_comp}}
\end{center}
\end{figure*}

The left panel of Fig.~\ref{vrmsconv_comp} compares the time-averaged root-mean-square velocity as a function of radius in the deep stellar interior for four simulations that include different extents of the radiative zone, but all exclude the near-surface layers.   In the portion of the radiative zone we consider, which spans approximately $0.1 < r/R < 0.42$, discrete large amplitude waves are destabilized.  These waves propagate in the angular direction and can survive for long times; similar waves are identified by \citet{alvan2015characterizing} as possible gravity waves.   Simulations Low4 and Low6, which include a large extent of the radiative zone, display waves of particularly large amplitude in the radiative zone.  However the amplitude of these waves does not appear to consistently increase with the extent of the radiative zone that is included in the spherical shell for simulations Low1-9.  In addition, waves of similarly large amplitude are not present in the radiative zone of simulations Hi2 or ExH2, which are higher resolution but otherwise identical to simulation Low4. The length of time of the simulation appears to be critical; the tens of convective turnover times of simulations Hi2 or ExH2 may not be a sufficient amount of time to observe the excitation of large amplitude waves.  We speculate that the large amplitude of these waves may be produced during intermittent events.   We also observe no clear correlation in our simulations between the amplitudes of the waves in the radiative zone and the amplitude of  $\langle v_{\mathsf{RMS}}\rangle_t$ in the lower convection zone.

Although the extent of the radiative zone included in the simulations in the left panel of Fig.~\ref{vrmsconv_comp} is different, the radial profiles of $\langle v_{\mathsf{RMS}}\rangle_t$ are remarkably similar in the convection zone.  
The bottom of the convection zone, indicated by a heavy vertical line, is calculated from the entropy profile produced by the one-dimensional Lyon stellar evolution code.
Above this boundary, $\langle v_{\mathsf{RMS}}\rangle_t$ smoothly rises to a maximum in each simulation.  
This maximum in the time-averaged RMS velocity lies approximately in the range $200$m/s to $300$m/s for these four simulations.    Throughout the rest of the convection zone, the time-averaged RMS velocity remains relatively flat.  The near-surface layers are omitted from these simulations and are not shown on this plot.   An area containing one standard deviation above and below the time-average for simulation Low2 is shaded.   Because the average values lie within a standard deviation of each other, we consider the dynamics of these four simulations to be indistinguishable.  We conclude that, as long as the boundary at the bottom of the convection zone is included, the convective dynamics are largely unaffected by the extent of the radiative zone included in the spherical shell.

The right panel of Fig.~\ref{vrmsconv_comp} shows the radial profile of $\langle v_{\mathsf{RMS}}\rangle_t$ for simulations Low7 and Low8.  These two simulations include the near-surface layers and are identical except that Low7 does not include the boundary at the bottom of the convection zone.   Eliminating this natural boundary clearly results in a significantly elevated profile of the time-averaged RMS velocity in the lower convection zone.   We observe that when some extent of the radiative zone is included in the simulation, the radiative zone appears to act as an energetic buffer.  Energy from convective motions that reach the radiative boundary may contribute to waves in the radiative zone.  In simulation Low7, where the radiative boundary is not included, all ballistic motions are simply reflected back into the convection zone.  Higher velocity structures in the convection zone result.   

We conjecture that this result could vary for different stellar models.  A peculiarity of our young sun model is that the energy flux changes with the radius.  Because Low7 and Low8 use boundary conditions consistent with the radial profile of energy flux in the young sun model, the energy flux through their lower boundaries is not identical.  Placing the boundaries in different physical zones could lead to different build-ups of energy.

There is approximately a factor of 5 difference between the time-averaged RMS velocity produced in our two-dimensional hydrodynamics simulations and the mixing-length-theory velocity used in the one-dimensional calculation.  This observation is not independent of the factor of 5 found in our comparison of convective turnover times in Section \ref{secturntime}.  The pressure scale-height is used to produce the convective turnover time from the RMS velocity; however the pressure scale-height is indistinguishable between one- and two-dimensional simulations over the times we examine.

\subsubsection{Coupling to the near-surface layers: impact on velocity}

\begin{figure*}[h]
\begin{center}
\resizebox{3.5in}{!}{\includegraphics{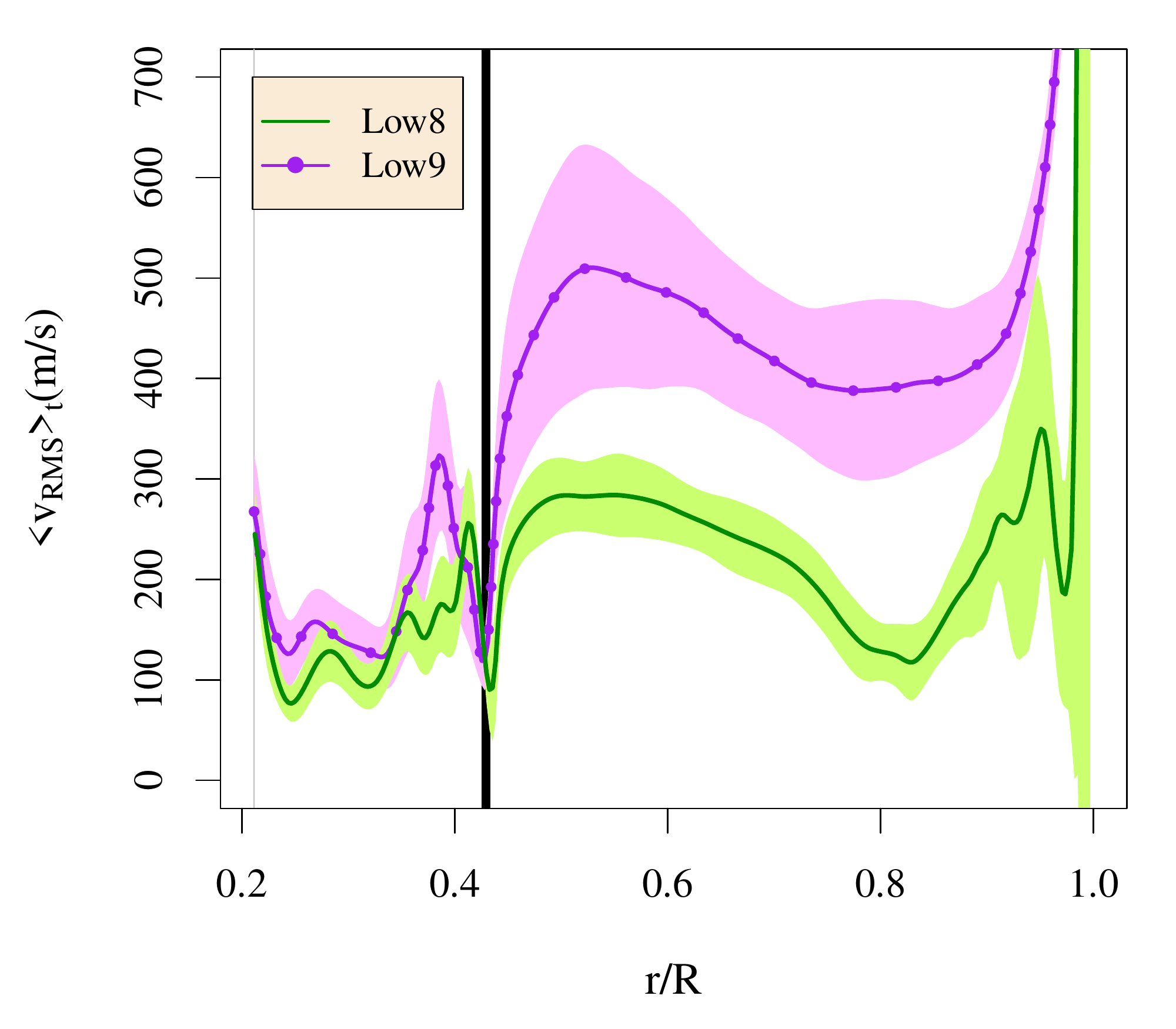}}\resizebox{3.5in}{!}{\includegraphics{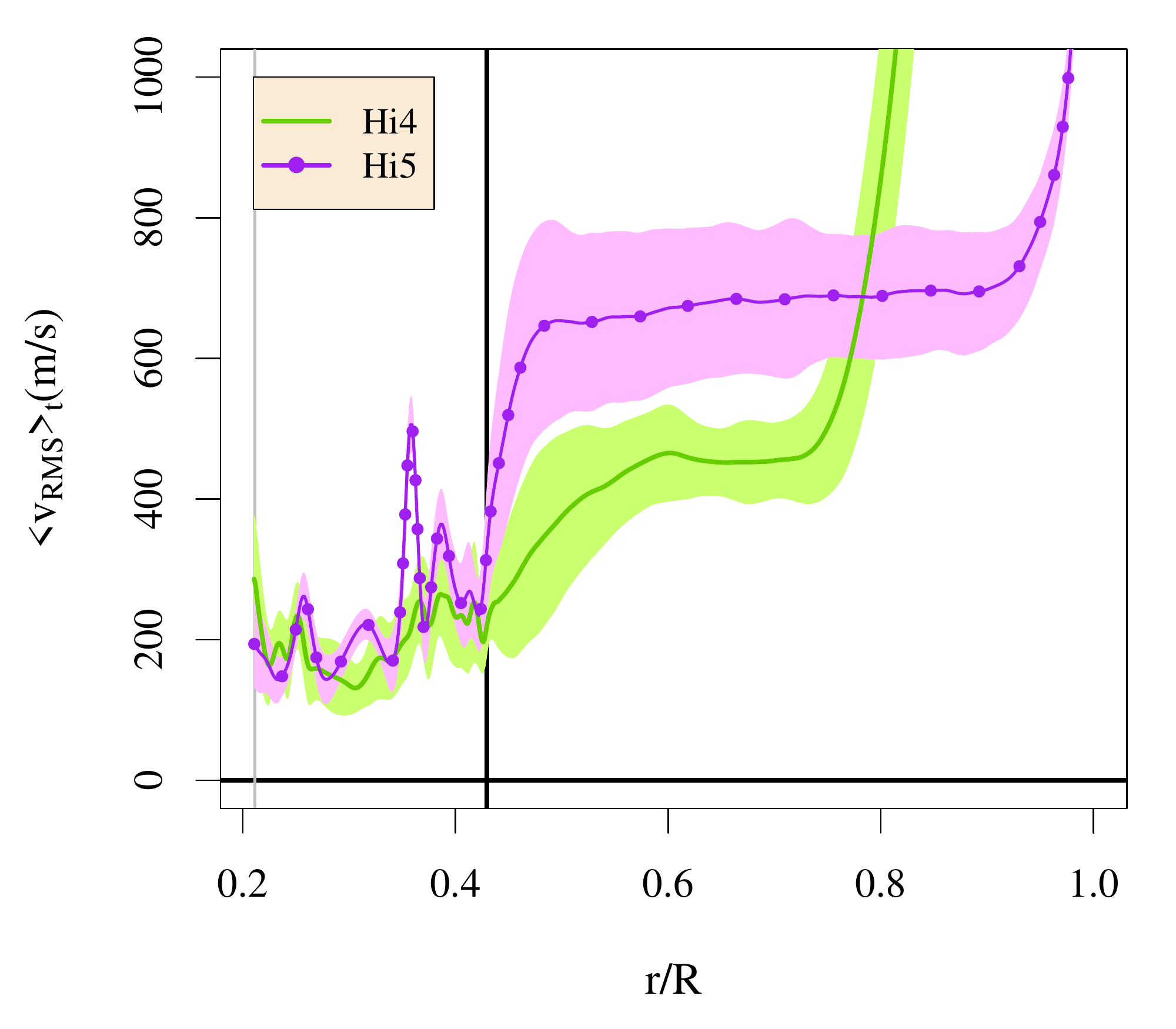}}
\caption{Time-averaged radial profile of RMS velocity in (left) simulations Low8 and Low9 which include the near-surface layers up to the surface at $r/R=1.0$, and (right) simulations Hi4 and Hi5.   The shaded areas indicate one standard deviation above and below each time-averaged line.  A heavy vertical black line marks the boundary between the radiative and convection zones, calculated from the radial profile of entropy and the Schwarzschild criterion.  Grey vertical lines mark the inner radial boundary of the spherical shell simulation volume for each simulation shown.
 \label{vrmssurf_comp}}
\end{center}
\end{figure*}

The left panel of Fig.~\ref{vrmssurf_comp} compares the amplitude of time-averaged RMS velocities in the stellar interior for two simulations that include the near-surface layers, Low8 and Low9.  These simulations use identical simulation volumes, but differ in how well the near-surface layers are resolved; simulation Low9 uses the spliced grid where grid spacing decreases in the near-surface layers, with the result that the grid spacing near the surface is approximately half the size of that in simulation Low8.  In addition, simulation Low9 is allowed to radiate heat dependent on the local temperature on the boundary, \emph{i.e.} in a non-spherically symmetric way, as discussed in Section \ref{secbc}.  The different treatment of the near-surface layers has a significant impact on convective velocities throughout the convection zone, in line with the predictions of \citet{spruit1997convection}.  The maximum of $\langle v_{\mathsf{RMS}}\rangle_t$ at the bottom of the convection zone is slightly less than twice as large in simulation Low9 as in Low8.  The shaded areas that indicate one standard deviation above and below the time-averages for Low8 and Low9 do not overlap.  Aside from the difference in velocity magnitude throughout the convection zone, the profiles of $\langle v_{\mathsf{RMS}}\rangle_t$ exhibit a similar shape in simulations Low8 and Low9, indicating that the radial structure of stellar convection is preserved.  Although distinctly different stars are produced by simulations Low8 and Low9, the approximate shape of the radial profile of $\langle v_{\mathsf{RMS}}\rangle_t$ in the upper convection zone is similar.  This trend of increasing RMS velocity with radius can also be compared to results produced for the current sun by the ASH, MURaM, and STAGGER codes \citep[see Figure 4 of][]{miesch2012amplitude}.

The right panel of Fig.~\ref{vrmssurf_comp} compares the radial profile of time-averaged RMS velocity in simulations Hi4 and Hi5.  Simulation Hi5 has the same treatment of the surface layers as simulation Low9, using a spliced grid to better resolve the temperature gradient in the near-surface layers in order to allow energy to radiate as black-body radiation.   Simulation Hi4 is identical to simulation Low8, but uses double the grid-size.  Although simulation Hi4 experiences a faster rise in RMS velocity in the upper convection zone than simulation Hi5, a larger magnitude of time-averaged RMS velocity is observed throughout the lower convection zone for simulation Hi5 than for simulation Hi4.  In comparison with the left panel of Fig.~\ref{vrmssurf_comp}, the simulations in the right panel show that, in the lower convection zone, a higher velocity is generally produced when the grid spacing is reduced.  This results because we use only numerical dissipation of the velocity, which is reduced with higher resolution.  Aside from this difference in amplitude, we find that properties of the flow in the lower convection zone produce a comparable form for the radial structure of the overshoot region, for the resolutions we are able to investigate.  This is reassuring for future efforts to numerically study overshooting at the lower boundary of the convection zone.   The differences observed in the upper convection zone are understandable because the treatment of the surface clearly impacts the character of surface convection.

\subsection{Overshooting layer width \label{sectionovershoot}}

Large-scale convective motions behave ballistically, terminating at a range of radial points near the boundary with the convectively stable radiative zone.
This situation can be classified as either convective overshooting or convective penetration, a distinction based on the impact that the convective flow has on the physics of the radiative zone \citep{brummell2002penetration,maederbook,zahn1991convective}.  
In our simulations, we use a realistic thermal diffusivity that is identical to that used in the one-dimensional stellar evolution calculation, and has not been artificially enhanced.  Because of this, our simulations are in the large P\'eclet number regime in the deep stellar interior, and penetrative convection can be theoretically expected at the lower boundary of the convection zone.   Simulations at large P\'eclet numbers in deep stellar interiors have been reported by \citet[][]{meakin2007turbulent,viallet2015toward}.  

An aspect of the dynamic structure of a star that results from the dynamic coupling across the boundary between the radiative zone and convection zones is the width of the overshooting layer.
 The width of the overshooting layer is sometimes called an overshooting length, and is related to an overshooting parameter in one-dimensional simulations \citep[e.g.][]{schroder1997critical,zhang2012turbulent,renzini1987some}.  
Overshooting establishes a layer between the stable radiative zone and the convection zone, where convective motions mix into a more quiescent fluid.  The physics of this layer has interesting implications for the stellar chemistry, stellar evolution, and magnetic field generation \citep{marik2002new,marques2006effect,tian2009numerical}. We do not address the physics of the overshooting layer in this work aside from evaluating its width.  The overshooting layer width that we calculate here is a convenient measure that can be benchmarked for simulations in different spherical shells; it is not the same as the overshooting length used in one-dimensional stellar evolution calculations.
  The width of the overshooting layer can only be defined from a long-time average of the dynamics, which quantifies the interaction between the convection and radiative zones.
  
\begin{figure*}
\begin{center}
\resizebox{3.5in}{!}{\includegraphics{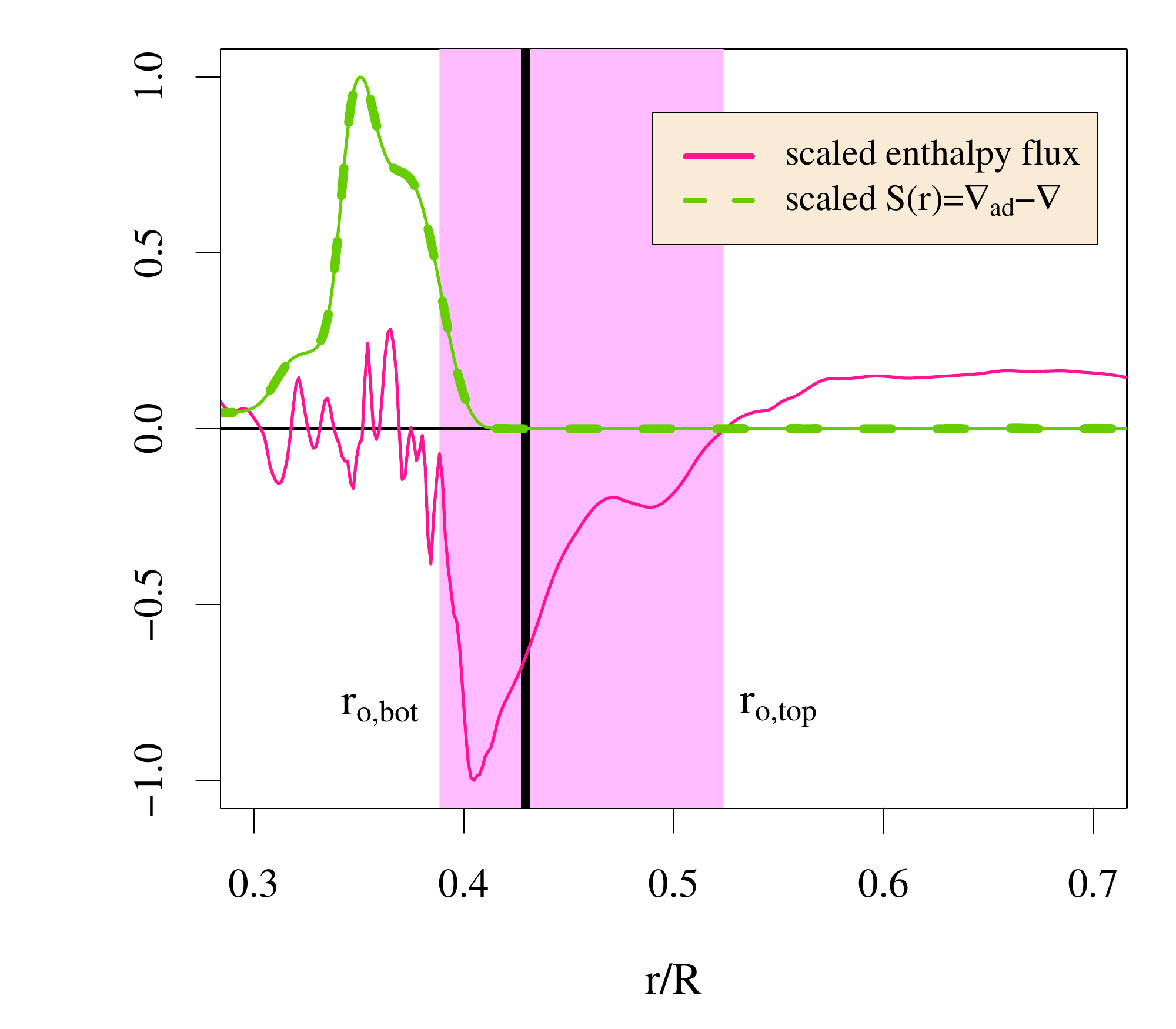}}\resizebox{3.5in}{!}{\includegraphics{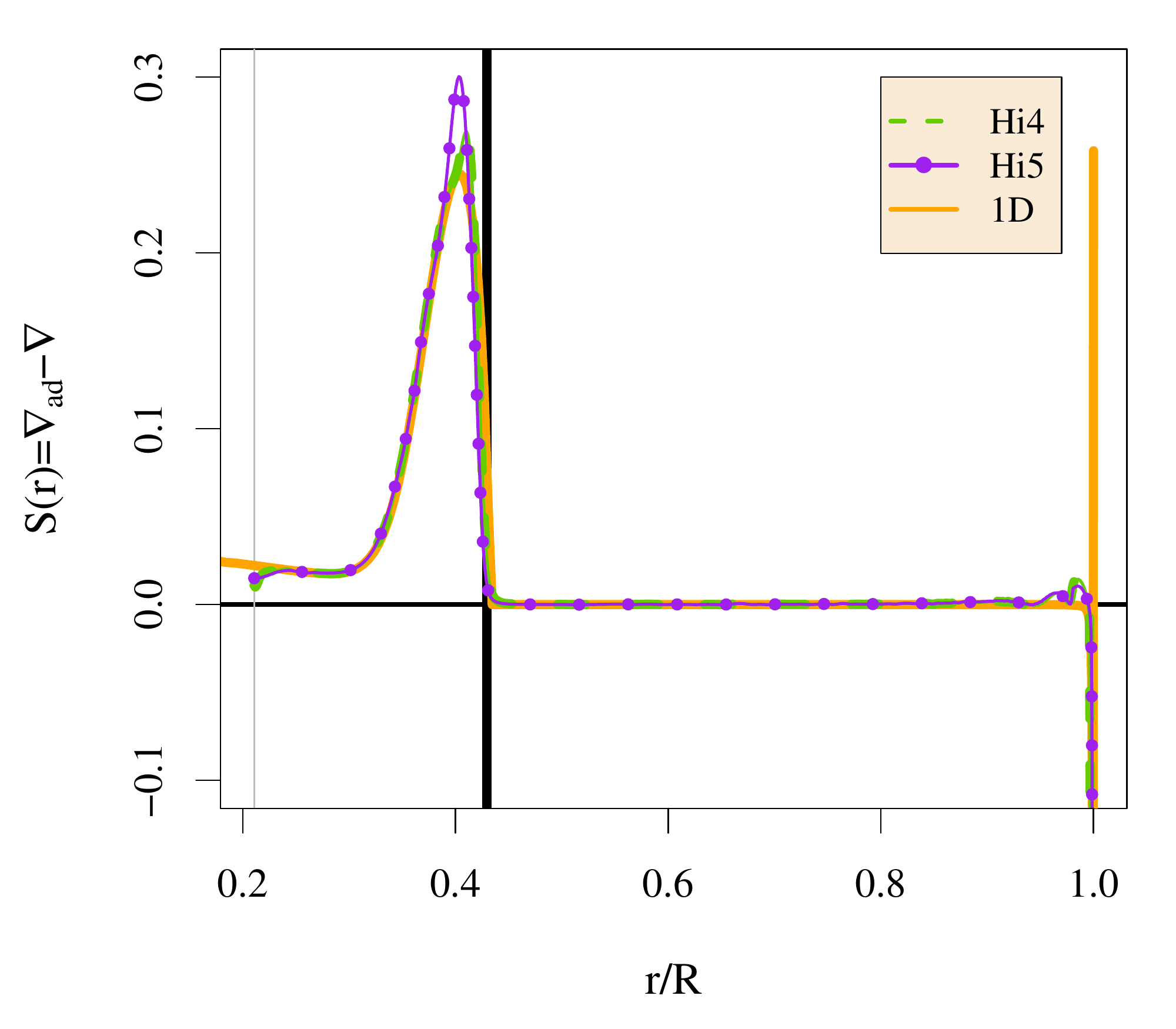}}
\caption{(Left) overshooting layer in simulation Hi1, represented by the shaded area between the radii of $r_{\mathsf{o,top}}$ and $r_{\mathsf{o,bot}}$.  The heavy vertical line shows the position of the boundary between the radiative and convection zones, indicated by the entropy profile produced by the one-dimensional stellar evolution calculation.  
The full line shows the enthalpy flux, scaled by the magnitude of the large negative peak.  The dashed line shows the Schwarzschild discriminant $S(r)$, scaled to appear on this graph.  (Right) the Schwarzschild discriminant in simulations Hi4 and Hi5, compared with the value obtained from the one-dimensional model of the young sun.
\label{figenflux}}
\end{center}
\end{figure*}

To evaluate the sensitivity of the overshooting layer to the spherical shell geometry and boundary conditions, we adopt one possible method for calculating a width for the overshooting layer described in \citet{brun2011modeling,browning2004simulations}.  We define the overshooting layer width:
\begin{eqnarray}\label{eqovershootingwidth}
w_{\mathsf{o}} = r_{\mathsf{o,top}} - r_{\mathsf{o,bot}} 
\end{eqnarray}
where $r_{\mathsf{o,top}}$ is the radius at the top of the overshooting layer, defined as the point where the time average of the enthalpy flux first changes sign.  In Table~\ref{simsuma}, $r_{\mathsf{o,top}}$ is listed for each simulation.  The bottom of the overshooting layer, $r_{\mathsf{o,bot}}$, is defined
as the radius in the radiative zone where the enthalpy flux becomes negligible.  In practice $r_{\mathsf{o,bot}}$ is also the point where waves in the radiative zone begin to
impact the radial profile of time-averaged enthalpy flux.  We define the enthalpy flux following \citet{freytag1996hydrodynamical}:
\begin{eqnarray}\label{defenthalpyflux}
F_{\mathrm{H}} = \langle H  \rho v_{\mathsf{r}} \rangle - \langle H \rangle \langle  \rho v_{\mathsf{r}} \rangle
\end{eqnarray}
where the enthalpy $H = e + P/\rho$ is standardly calculated in terms of the pressure  $P$, density  $\rho$, and internal energy $e$.  The second term in eq.~\ref{defenthalpyflux} subtracts the effect of any bulk mass flow, which is typically small.
The enthalpy flux in the region surrounding the overshooting layer is illustrated in the left panel of Fig.~\ref{figenflux} for simulation Hi1. Both $r_{\mathsf{o.top}}$ and $r_{\mathsf{o,bot}}$ are labeled in this figure.  
The Schwarzschild discriminant $S(r) = \nabla_{\mathsf{ad}} - \nabla$ is also shown  in  Fig.~\ref{figenflux} for simulation Hi1, as well as for simulations Hi4 and Hi5.  In the Schwarzschild discriminant, the so-called adiabatic gradient $\nabla_{\mathsf{ad}}= \left| \partial \log{T}/\partial \log{P} \right|_{\mathsf{ad}}$ is calculated using the equation of state.  The local gradient $\nabla= \left|\partial \log{T}/\partial \log{P} \right|$ is calculated from our hydrodynamic simulations and appropriately averaged.  The Schwarzschild criterion amounts to the statement that the Schwarzschild discriminant, $S(r)$, must be greater than zero for stability against convection \citep{osterbrock1958structure,lebovitz1965schwarzschild}.  We see small deviations from the one-dimensional profile in the Schwarzschild discriminant in these figures.
We do not expect that these small deviations are due to convective penetration because of the limited time that the simulations are observed in comparison with the time scale for thermal evolution at this depth.  The small differences possibly stem from two-dimensional fluid effects and the low resolution of the fluid simulations compared to the one-dimensional stellar evolution calculation.

\begin{figure*}
\begin{center}
\resizebox{3.5in}{!}{\includegraphics{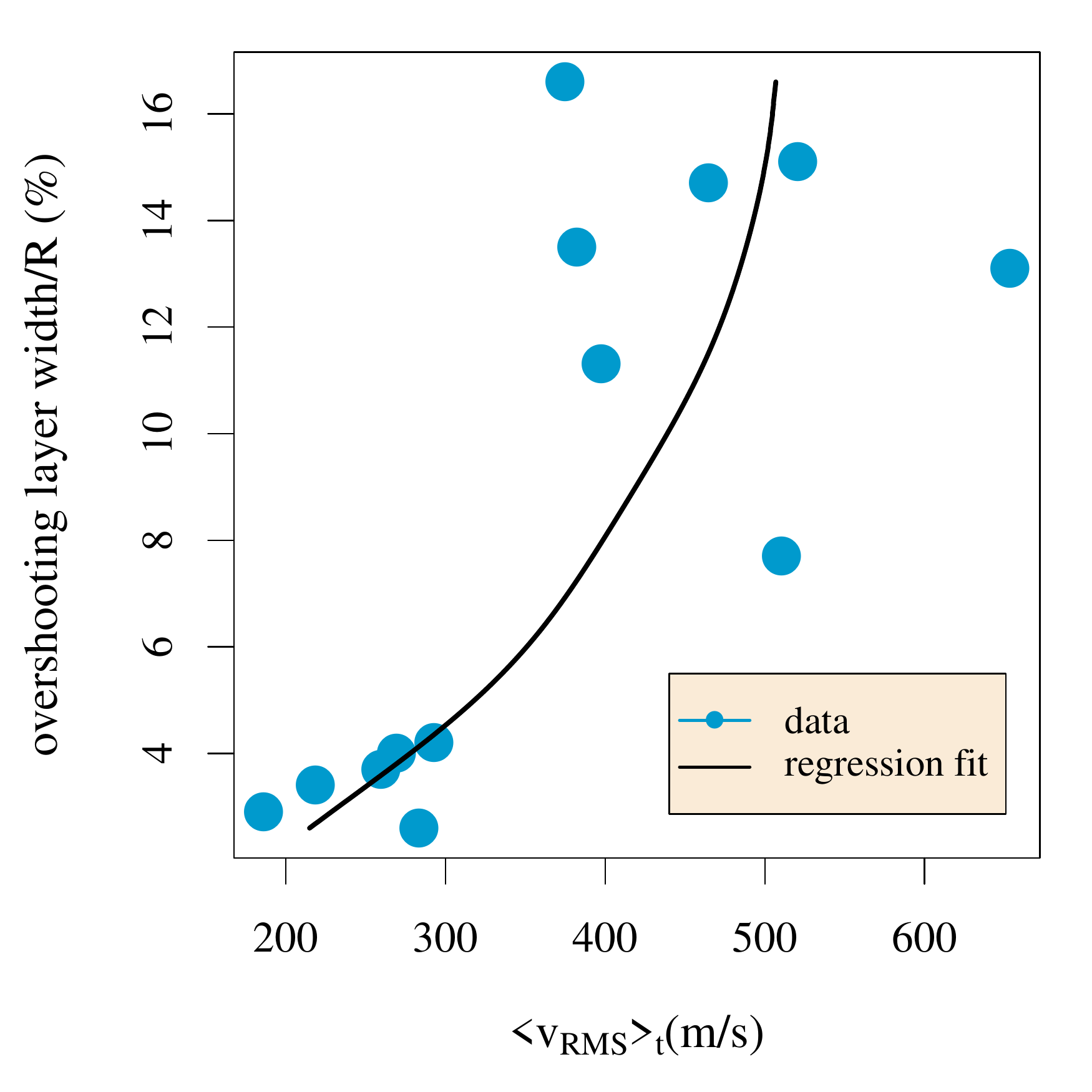}} 
\caption{Overshooting layer width defined in eq.~\eqref{eqovershootingwidth} \emph{vs} the maximum of the time-averaged root-mean-square velocity above the radiative zone boundary in the overshooting layer.   A local polynomial fit to the data is also shown.
\label{figwidthcorr}}
\end{center}
\end{figure*}

The overshooting layer width, $w_{\mathsf{o}}$, for each simulation is given in Table~\ref{simsuma} as a percentage of the total stellar radius $R$.  For simulations Low1-9, the overshooting layer width is typically about 4\% of the young sun's radius, or $0.21 h_p$ at the boundary to the radiative zone; for the higher resolution simulations Hi1-5, it is approximately 14\% of the young sun's radius, or $0.76 h_p$.   Intuitively, the width of the overshooting layer is linked to the velocity amplitude.  The difference in the overshooting layer width that we observe in simulations of different resolution clearly reflects the different velocity of convection rolls in the convection zone, and indicates a higher level of interaction between these zones when higher velocities are present.  This relationship is shown in Fig.~\ref{figwidthcorr}; the degree of certainty for the largest overshooting layer widths recorded in this figure are low, because they correspond to high-resolution simulations that were performed for comparatively short times.   Beyond this link to local velocity amplitudes, the overshooting layer width appears to be independent of the spherical shell geometry and boundary conditions.

When the overshooting layer width is larger, its growth is not centered around the boundary where the Schwarzschild criterion indicates stability. It begins substantially higher in the convection zone and expands proportionately less into the radiative zone. For Low1-9 we measure $r_{\mathsf{o,bot}}/R \approx 0.41$, while for simulations Hi1-5 this value is $r_{\mathsf{o,bot}}/R \approx 0.39$.  We observe that the inner radial boundary of simulation Low1 is identical to $r_{\mathsf{o,bot}}$; this likely impacts the unusually large overshooting layer width in this simulation.

\section{Summary and Discussion}

We have examined how the extent of a two-dimensional spherical shell, and therefore the boundary conditions imposed on an interior convection zone, affect the characteristics of compressible convection.   A spherical shell that extends deeper into the central radiative zone is used to analyze the impact of a physically-motivated open boundary condition on the inner radius of the convection zone.   A spherical shell that extends into the near-surface layers is used to analyze the impact of an open boundary condition on the outer radius of the convection zone.   In the non-local theory of stellar convection, coupling between different zones of a star is expected to strongly influence stellar dynamics.

The results of our numerical experiments show that this kind of coupling indeed can impact the amplitude of the convective velocities that evolve throughout the convection zone.   The inclusion of the radiative zone as an inner radiative boundary on the convection zone produces a lower velocity of convection rolls throughout the convection zone.  The presence of small-scale convection in modeled near-surface layers also impacts the amplitude of convection rolls throughout the convection zone.  
The combined effect of spherical shell geometry and boundary treatment can result in a difference greater than a factor of two in our velocity data. 
Thus even though the flows in our modeled near-surface layers are not well resolved, we do observe a non-local effect in agreement with \citet{spruit1997convection}.  To resolve the flows in the near-surface layers well and allow more accurate coupling with the convection zone, adaptive mesh refinement is a promising tactic; this is a development direction that we are studying.  Along with an increase in the amplitude of convective velocities, a predictable increase in the width of the overshooting layer, and decrease in the convective turnover time is generally observed.

The salient aspect of these results for our ongoing studies of three-dimensional convective overshoot is that, although coupling between different zones changes the amplitude of convective velocities, the radial structure of the convective dynamics in the young sun maintains a strikingly similar profile,  in particular in the overshoot region.  Thus a simulation that comprises accurate modeling of the near-surface layers of the young sun will be able to produce results for deep convective processes physically similar to simpler models.  Those results will be relevant to a lower velocity regime, but can still contribute in critical ways to the 321D link.  In general the results from simulations also may be more realistic if the region of interest is far from the surface, such as fluid and chemical mixing, shear-flows in a deep interior, or overshooting at the lower boundary of the convection zone.

This work has focused on analyzing the spherical shell geometry and boundary conditions.  In summary, we obtain two helpful observations: (1) that coupling between different zones in the young sun model changes the amplitude of dynamic quantities, but (2) the dynamic structure of the star is not affected. 
The simulations we analyze in this work are two-dimensional and focus exclusively on large-scale stellar flows.  We therefore refrain from 
 physical conclusions broader than the impact of the spherical shell geometry and boundary conditions.  A focused, more complete study of convection and convective overshooting in the young sun, including three-dimensional simulations, is planned based on these results.  A broader physical discussion of convection may await those calculations.

\FloatBarrier
\begin{acknowledgements}
The research leading to these results has received funding from the European Research
Council under the European Union's Seventh Framework (FP7/2007-2013)/ERC grant
agreement no. 320478.
\\
M. Viallet is funded by the European Research Council though grant ERC-AdG No. 341157-COCO2CASA.
\\
This work used the DiRAC Complexity system, operated by the University of Leicester IT Services, which forms part of the STFC DiRAC HPC Facility (www.dirac.ac.uk). This equipment is funded by BIS National E-Infrastructure capital grant ST/K000373/1 and STFC DiRAC Operations grant ST/K0003259/1. DiRAC is part of the National E-Infrastructure.
\\
This work also used the University of Exeter Supercomputer, a DiRAC Facility jointly funded by STFC, the Large Facilities Capital Fund of BIS and the University of Exeter.
\end{acknowledgements}

\bibliographystyle{aa}
\bibpunct{(}{)}{;}{a}{}{,}
\bibliography{music}

\end{document}